\documentclass{raa}

\renewcommand{\RAAlogo}{}
\usepackage{graphicx,times}             
\usepackage{natbib}
\usepackage{amssymb,amsmath}
\bibpunct{(}{)}{;}{a}{}{,}

\usepackage{fix-cm}   

\usepackage[paperwidth=210mm, paperheight=280mm, top=2.5cm, bottom=2.5cm, left=2cm, right=2cm]{geometry}
\usepackage{subcaption}
\usepackage{bm}
\usepackage{newtxtext,newtxmath}
\usepackage{hyperref}

\begin{document}

\title{Atmospheric Dynamics of Asymmetrically Magnetized Hot Jupiter}

   \volnopage{Vol.0 (202x) No.0, 000--000}      
   \setcounter{page}{1}          
   
\author{Miaoyin Tang\inst{1} \and Cong Yu\inst{1,2,3,4}}

\institute{
  School of Physics and Astronomy, Sun Yat-sen University, Zhuhai 519082, China; \\
  {\it yucong@mail.sysu.edu.cn}
  \and
  CSST Science Center for the Guangdong-Hong Kong-Macau Greater Bay Area, Zhuhai 519082, China
  \and
  State Key Laboratory of Lunar and Planetary Sciences, Macau University of Science and Technology, Macau, China
  \and
  International Centre of Supernovae, Yunnan Key Laboratory, Kunming 650216, China
}

\vs\no
{\small Received 2026 March 30; accepted 2026 May 26}

\abstract{ We investigate the influence of asymmetric magnetic fields on the atmospheric circulation of hot Jupiters based on a Shallow-Water Magnetohydrodynamic (SWMHD) model. The Shallow-Water Hydrodynamic (SWHD) models predict eastward equatorial jets and hotspot offsets, while some observations have revealed westward hotspots, suggesting that magnetic fields may play an important role. We incorporate asymmetric magnetic fields between hemispheres, and analyze their effects through linear perturbation analysis and numerical calculations. Our results indicate that strong magnetic fields play a dominant role in momentum transport. Asymmetric magnetic field configurations lead to hemispheric temperature contrasts, with the dayside temperature maxima in the hemisphere of stronger magnetic field located closer to the equator. With the magnetic field fixed in one hemisphere, the equatorial hotspots shift westward then eastward as the other hemisphere’s field strengthens, exhibiting a pronounced westward offset only at moderate field strengths and weak hemispheric asymmetry. These findings highlight the significance of magnetic field geometry in explaining observed atmospheric dynamics and hotspot variability in hot Jupiters.
}
    \authorrunning{ Tang \& Yu }     
   \titlerunning{Atmospheric Dynamics of Asymmetrically Magnetized Hot Jupiter }  

\maketitle

\section{Introduction} \label{section 1}
Hot Jupiters were the first detected exoplanets \citep{1995Natur.378..355M}. As gas giants with masses and sizes comparable to that of Jupiter in the Solar System, they orbit their host stars at an extremely close distance. Subject to tidal locking, hot Jupiters have a permanent dayside and a permanent nightside. Their outer atmospheres are exposed to intense stellar radiation, resulting in high temperatures. Therefore, studying the atmospheres of these planets is key to understanding atmospheric dynamics, planetary evolution\citep{2020RAA....20...99Z,2024RAA....24f5022X}.

Current analyses of observational data can constrain the three-dimensional temperature structures of several hot Jupiters and indicate intense atmospheric circulation on these objects\citep{2009ApJ...690..822K,2016ApJ...821...16K,2024NatAs...8..879B,2024ApJS..270...34C,2024PASP..136h4403W,2025ApJ...986...63W}. In the absence of consideration for atmospheric advection within the hot Jupiter, its hotspot (i.e., the point of maximum luminosity) should be located at the substellar point directly irradiated by the host star.
Generally, observational measurements of hot Jupiters find that these planets have equatorial temperature maxima (hotspots) located eastward of their substellar points \citep{2006Sci...314..623H,2007Natur.447..183K,2008ApJ...686.1341C,2009ApJ...690L.114S,2010ApJ...723.1436C,2014Sci...346..838S}. 
The research of \cite{2011ApJ...738...71S}  showed that SWHD models of hot Jupiters will always produce eastward equatorial hotspots. Tidally locked short-period planets exhibit a mechanism that drives equatorial superrotation \citep{2018AREPS..46..175R,2020SSRv..216..139S,2022ApJ...927...38I,2023PSJ.....4..106L}.The strong day-night temperature difference forms coupled Rossby-Kelvin waves, transporting eddy momentum toward the equator and generating equatorial superrotation \citep{2011ApJ...738...71S,2020SSRv..216...87I,2021ApJ...907...28W}. 
However, recent observations show that some hot Jupiters exhibit westward hotspots \citep{2018NatAs...2..220D,2019MNRAS.489.1995B,2021MNRAS.504.3316B}. These observations suggest the existence of a mechanism that can also drive westward equatorial winds and there are three main explanations at present: non-synchronous rotation \citep{2014ApJ...790...79R}, asymmetric cloud distribution caused by clouds tending to form in regions with lower atmospheric temperatures \citep{2015PNAS..11213461G,2016ApJ...828...22P,2019ApJ...872....1R} and magnetic drag \citep{2010ApJ...724..313P,2022AJ....163...35B}. The toroidal magnetic field of hot Jupiters intensifies rapidly with increasing temperature, producing a drag effect that significantly impacts the atmospheric circulation of hot Jupiters. \cite{2019ApJ...872L..27H} used shallow-water magnetohydrodynamic (SWMHD) simulations to show that a large planetary magnetic field can explain the westward atmospheric winds and westward hotspots on hot Jupiters. Magnetic effects can slow down the equatorial jets and even reverse the flow toward the west
\citep{2013ApJ...764..103R,2014ApJ...794..132R,2021ApJ...916L...8H,2021ApJ...922..176H,2023ApJ...959...41H}.
\cite{2024ApJ...969L..32C} first detected a significant latitudinal (north-south) hotspot offset in the dayside atmosphere of WASP-43b, with the hottest point not located at equator but shifted toward the southern hemisphere, potentially due to the influence of a very strong magnetic field. 

The magnetic field of hot Jupiters is influenced by various factors. Notably, north–south hemispheric asymmetry of magnetic fields can arise from heterogeneous interior dynamics, which have been observed in the asymmetric magnetic field of Jupiter \citep{2018AGUFM.P23A..07C,2018NatAs...2..220D} and may exist on hot Jupiters. Magnetohydrodynamic simulations of a hot Jupiter atmosphere show that a sufficiently strong background magnetic field initiates the subcritical dynamo process, generating intense local magnetic fields \citep{2025A&A...699A.339B}. The day-night temperature difference and the inclusion of conductivity variations on hot Jupiters can further reshape the magnetic field structure. When magnetic flux is swept from the strongly coupled dayside to the dissipative nightside \citep{2017ApJ...841L..26R}, it leads to complex field topology and  generates a highly asymmetric zonal magnetic field. Such hemispheric magnetic asymmetry can have a substantial impact on the atmospheric circulation of hot Jupiters.

Based on many pioneering investigations that have been published
\citep{2008ApJ...685.1324S,2009ApJ...699..564S,2009ApJ...703.1819H,2010ApJ...716..144T,2010ApJ...724..313P,2011ApJ...738...71S}, here we present the first implementation of asymmetric magnetic fields in an atmospheric circulation model to examine their influence. First, we choose the SWMHD model and then use linear perturbation analysis to reproduce both eastward hotspot offset in hydrodynamic cases and westward hotspot offset with equatorially antisymmetric magnetic field. Then, we add a north-south asymmetric magnetic field to investigate its effect on the atmospheric circulation of hot Jupiters. Finally, we introduce the zonal (east-west) acceleration of the zonal-mean flow to investigate the role of magnetic fields in momentum transport.

\section{SWMHD Model} 
\label{section 2}
We adopt the SWMHD model, which is the magnetohydrodynamics analog  of the shallow-water hydrodynamics models  widely used in hot Jupiter’s atmosphere \citep{2011ApJ...738...71S, 2000ApJ...544L..79G}. The SWMHD model has two constant-density fluid layers, the shallow upper layer represents the meteorologically active upper atmosphere, the infinitely deep lower layer represents the quiescent deep atmosphere and the interior of a hot Jupiter and the interface between the two layers is a material surface with no magnetic flux passing through it. The active layer's thickness $H$ (pressure scale height) is much smaller than the typical active-layer horizontal scale $L$, the system approaches magneto-hydrostatic balance and the variables of the SWMHD equations become independent of the vertical coordinate. For a local Cartesian system in the equatorial $\beta$-plane approximation, the shallow active upper layer is governed by the following equations:
\begin{align}
&\frac{d\textbf{V}}{dt} + g\nabla h + \beta y (\textbf{k}\times\textbf{V})  = \textbf{R}+ (\textbf{B}\cdot\nabla)\textbf{B}-\frac{\textbf{V}}{\tau_{drag}},
\tag{2-1}
\label{2-1}\\
&\frac{\partial h}{\partial t}+\nabla \cdot(\textbf{V}h)= \frac{h_{eq}(x,y)-h}{\tau_{rad}}\equiv Q, 
\tag{2-2}
\label{2-2}\\
&\frac{\partial A}{\partial t}+\textbf{V} \cdot \nabla A=\eta \left(\nabla^2A-\frac{1}{h}\nabla h\cdot\nabla A\right).  
\tag{2-3}
\label{2-3}
\end{align}
They represent momentum equation, continuity equation, and magnetic induction equation
\citep{2025JFM..1009A..50G}. The magnetic flux function $A$ is defined by $h\textbf{B}=\nabla \times A\hat{z}\equiv\left(\partial_yA,-\partial_xA,0\right)$, which automatically satisfies the SWMHD divergence-free condition $\nabla \cdot \left(h\textbf{B}\right)=0$. 
The magnetic field $\boldsymbol{B}\left(x,y,t\right) \equiv \left(B_x, B_y\right)$ denotes the horizontal magnetic field. It is actually related to the Alfvén wave speed $V_A = B_0/\sqrt{4\pi\rho}$ (the mass density $\rho$ is constant), so its unit is velocity. The vertical component of the magnetic field is neglected here for simplicity, as our focus is on the modulation of the horizontal circulation by the horizontal magnetic field. The vertical component will be incorporated in future three-dimensional models.

The horizontal gradient, Laplace operator and material derivative are defined by $\nabla \equiv (\partial_x, \partial_y)$, $\nabla^2=(\partial^2/\partial x^2+\partial^2/\partial y^2)$ and $d/dt=\partial/\partial t+\textbf{V} \cdot \nabla$. We use $\beta$-plane approximation to allow Coriolis parameter $f$ of the Coriolis force $-\Vec{f} \times \Vec{v}$ to have a functional dependence on $y$: $f=f_0 +\beta y$, where we have $f_0=2\Omega \cos \theta$, $\beta=2\Omega \sin\theta /R$, $\theta$ is polar angle, $\Omega$ is the angular frequency and $\textbf{k}$ is the upward unit vector. Near the equator, $f_0\approx0$, and $f$ can be approximately expressed as a linear function: $f\approx\beta y$. $h_{eq}$ represents the radiative-equilibrium height over the radiative timescale $\tau_{rad}$. The $\tau_{drag}$ term represents Rayleigh drag timescale, parameterizing the combined dissipative effects of frictional drag, turbulent dissipation, and other processes on the atmospheric fluid motion. 
$\eta$ is the magnetic diffusivity which behaves like molecular viscosity.

The term, ${\bf R}$, in equation \eqref{2-1}, represents the effect of Newtonian cooling term $Q$ on the momentum advection from the lower layer to the upper layer:
\begin{equation}
\textbf{R}(x,y,t)=
\begin{cases}
-\frac{Q\textbf{V}}{h},&Q>0;\\
0,&Q<0.
\end{cases}
\tag{2-4}
\label{2-4}
\end{equation} 
The magnetic field profile is expected to be equatorially antisymmetric  \citep{2014ApJ...794..132R}. So the background flux function $A_0$ can be represented in this form \citep{2019ApJ...872L..27H}:
\begin{equation}
A_0\left(y\right)=-e^{1/2}HV_AL_me^{-y^2/2L_m^2}.
\tag{2-5}
\label{2-5}
\end{equation}
The Alfvén speed $V_A$ determines the strength of the background magnetic field and the latitudinal decay length of the magnetic field $L_m =  \left(\pi L_{\rm eq}\right)/2$, where $L_{\rm eq}=(\sqrt{gH}/\beta)^{1/2}$ is defined as the equatorial Rossby deformation radius.

To enable the equations to describe an asymmetric magnetic field, we choose to introduce coefficients into the equations. 
\begin{equation}
A_0\left(y\right)=
\begin{cases}
-Y_1HV_AL_me^{1/2} e^{-y^2/(2L_m^2)},&y\leq0;\\
-Y_2HV_AL_me^{1/2} e^{-y^2/(2L_m^2)},&y\geq0.
\end{cases}
\tag{2-6}
\label{2-6}
\end{equation}
By changing the coefficients$\left(Y_1,Y_2\right)$ corresponding to the northern and southern hemispheres (y greater than or less than 0), we can obtain a magnetic field that is asymmetric across the hemispheres.

\section{Linear Analysis} \label{section 3}
We perform linear analysis of equations \eqref{2-1}-\eqref{2-3}, considering small perturbations to the physical quantities:
\begin{equation}
\begin{aligned}
V_x&=V_{x0}+u \ , \\
V_y&=V_{y0}+v \ , \\
h  &=H+\delta h \ , \\
B_x&=B_{x0}+b_x \ , \\
B_y&=B_{y0}+b_y \ ,\\
A&=A_0+a.
\end{aligned}
\tag{3-1}
\label{3-1}
\end{equation}
The background velocity and the  magnetic field in the $y$-direction are set to zero, $V_{x0}=V_{y0}=B_{y0}=0$. The magnetic field in the x-direction can be expressed as: $B_{x0}=\partial_y A_0/H$. We only retain the first-order terms and consider steady-state flows to get linearized equations:
\begin{equation}
\begin{aligned}
g\frac{\partial \delta h}{\partial x}-\beta yv+\frac{u}{\tau_{drag}}+\frac{1}{H^3}\left(\frac{\partial A_0}{\partial y}\right)^2\frac{\partial \delta h}{\partial x}-\frac{1}{H^2}\frac{\partial A_0}{\partial y}\frac{\partial^2 a}{\partial x\partial y}+\frac{1}{H^2}\frac{\partial^2A_0}{\partial y^2}\frac{\partial a}{\partial x}= 0,
\end{aligned} 
\tag{3-2}
\label{3-2}
\end{equation}
\begin{equation}
g\frac{\partial \delta h}{\partial y}+\beta yu+\frac{1}{H^2}\frac{\partial A_0}{\partial y}\frac{\partial^2a}{\partial x^2}+\frac{v}{\tau_{\rm drag}}=0  , 
\tag{3-3}
\label{3-3}
\end{equation}
\begin{equation}
H\left( \frac{\partial u}{\partial x}+\frac{\partial v}{\partial y}\right)+\frac{\delta h}{\tau_{rad}} = S(x,y) , 
\tag{3-4}
\label{3-4}
\end{equation}
\begin{equation}
\begin{aligned}
v\frac{\partial A_0}{\partial y}+\eta\frac{1}{H}\frac{\partial \delta h}{\partial x}\frac{\partial A_0}{\partial x}+\eta\frac{1}{H}\frac{\partial \delta h}{\partial y}\frac{\partial A_0}{\partial y}-\eta\left(\frac{\partial^2}{\partial x^2}+\frac{\partial^2}{\partial y^2}\right)a=0.  
\end{aligned}
\tag{3-5}
\label{3-5}
\end{equation}
Then we nondimensionalize equations \eqref{3-2}-\eqref{3-5} with length scale $L = (\sqrt{gH}/\beta)^{1/2}$ (the equatorial Rossby deformation radius), velocity scale $U = \sqrt{gH}$ (the gravity wave
speed), and timescale $T = L/U = (\sqrt{gH}\beta)^{-1/2}$ (the time for a gravity wave to cross a deformation radius in the shallow-water system). The thickness, the drag and thermal time, the forcing are nondimensionalized with $H$, $T$, and $H/T$. The magnetic field (in the units of velocity) and the flux function are nondimensionalized with $U$, $HUL$ . Define the magnetic Reynolds number $R_B = \frac{UL}{\eta}$, when $R_B \geq 1$, magnetic dynamo may operate \citep{2017eatc.book.....H}.
\begin{equation}
\begin{aligned}
\frac{\partial \delta h}{\partial x}-yv+\frac{u}{\tau_{drag}}+\left(\frac{\partial A_0}{\partial y}\right)^2\frac{\partial \delta h}{\partial x}-\frac{\partial A_0}{\partial y}\frac{\partial^2 a}{\partial x\partial y}+\frac{\partial^2A_0}{\partial y^2}\frac{\partial a}{\partial x}=0,\\
\frac{\partial \delta h}{\partial y}+yu+\frac{\partial A_0}{\partial y}\frac{\partial^2a}{\partial x^2}+\frac{v}{\tau_{drag}}=0,\\
\frac{\delta h}{\tau_{rad}}+\left(\frac{\partial u}{\partial x}+\frac{\partial v}{\partial y}\right)=S(x,y),\\
v\frac{\partial A_0}{\partial y}+\frac{1}{R_B} \frac{\partial \delta h}{\partial y}\frac{\partial A_0}{\partial y}-\frac{1}{R_B}\left(\frac{\partial^2}{\partial x^2}+\frac{\partial^2}{\partial y^2}\right)a=0.
\end{aligned}
\tag{3-6}
\label{3-6}
\end{equation}
Assuming that all physical quantities in $x$-direction are in the form of sine waves, satisfying the periodic boundary condition on the sphere $S(x,y) \propto e^{ik_x x}$, where $k_x$= $0.5$ is dimensionless zonal wavenumber. All physical quantities in $y$-direction are decomposed using the parabolic cylindrical equation: 
\begin{equation}
\psi_n\left(y\right)=A_n\exp\left(-\frac{y^2}{2}\right)H_n(y) \ , 
\tag{3-7}
\label{3-7}
\end{equation}
where $A_n=\pi^{-1/4}\left(2^n\cdot n !\right)^{-1/2}$ is the normalization constant and $H_n(y)$ is the Hermitian polynomial:
\begin{equation}
\begin{aligned}
H_n(y)&=(2y)^n-n(n-1)(2y)^{n-2}+\ldots
\\&+(-1)^{[n/2]}\frac{n!}{[n/2]}(2y)^{n-2[n/2]}, \\
H_n(y)&=2yH_{n-1}(y)-2(n-1)H_{n-2}(y), \\
H'_n(y)&=2nH_{n-1}(y).
\end{aligned}
\tag{3-8}
\label{3-8}
\end{equation}
The first- and second-order derivative relationships can be derived:
\begin{equation}
\begin{aligned}
&\frac{d \psi_n\left(y\right)}{dy}=-y\psi_n\left(y\right)+\left(2n\right)^{\frac{1}{2}} \psi_{n-1}\left(y\right),\\
&\frac{d^2 \psi_n\left(y\right)}{dy^2}=-\left(y^2-2n-1\right)\psi_n\left(y\right).
\end{aligned}
\tag{3-9}
\label{3-9}
\end{equation}
The boundary conditions in $y$-direction are that $u,v,h,A\to 0$ when $y\to \pm\infty$. 

\section{Numerical Solutions} \label{section 4}
\subsection{SWMHD Model Solutions with Asymmetric Magnetic Field}
To obtain concrete results from the dimensionless linearized equations \eqref{3-6}, we adopted the planetary parameters of a typical hot Jupiter. The nondimensional radiative and drag times are set to $\tau_{rad} = \tau_{drag} = 5$. The magnetic Reynolds number $R_B \approx 10^{3}$ is sufficiently large for magnetic fields to influence the atmospheric dynamics \citep{2022MNRAS.517.3113D}. 
\begin{figure*}[htbp]
    \centering
    \begin{subfigure}[b]{0.33\linewidth}
        \centering
        \includegraphics[width=\linewidth]{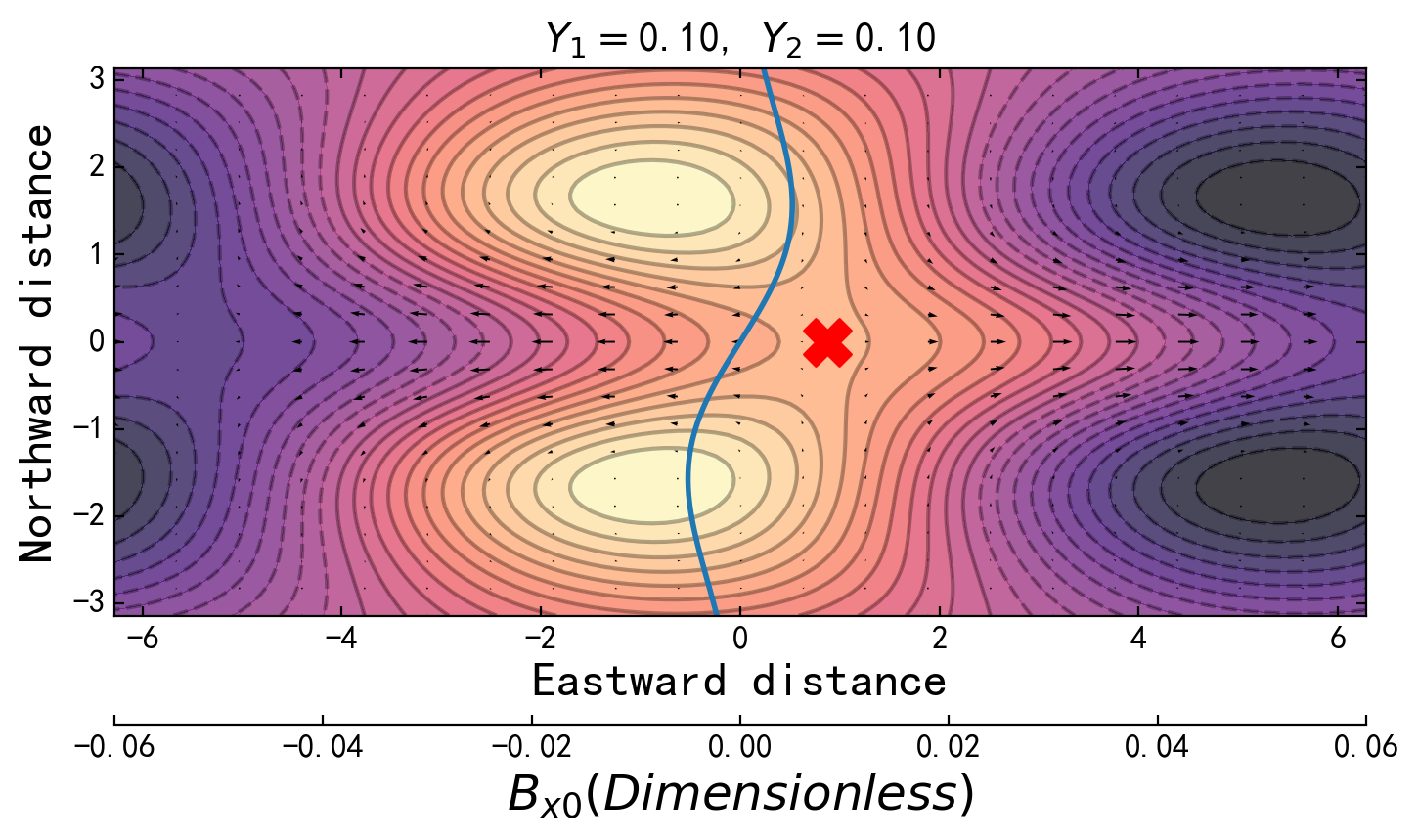}
    \end{subfigure}
    \begin{subfigure}[b]{0.33\linewidth}
        \centering
        \includegraphics[width=\linewidth]{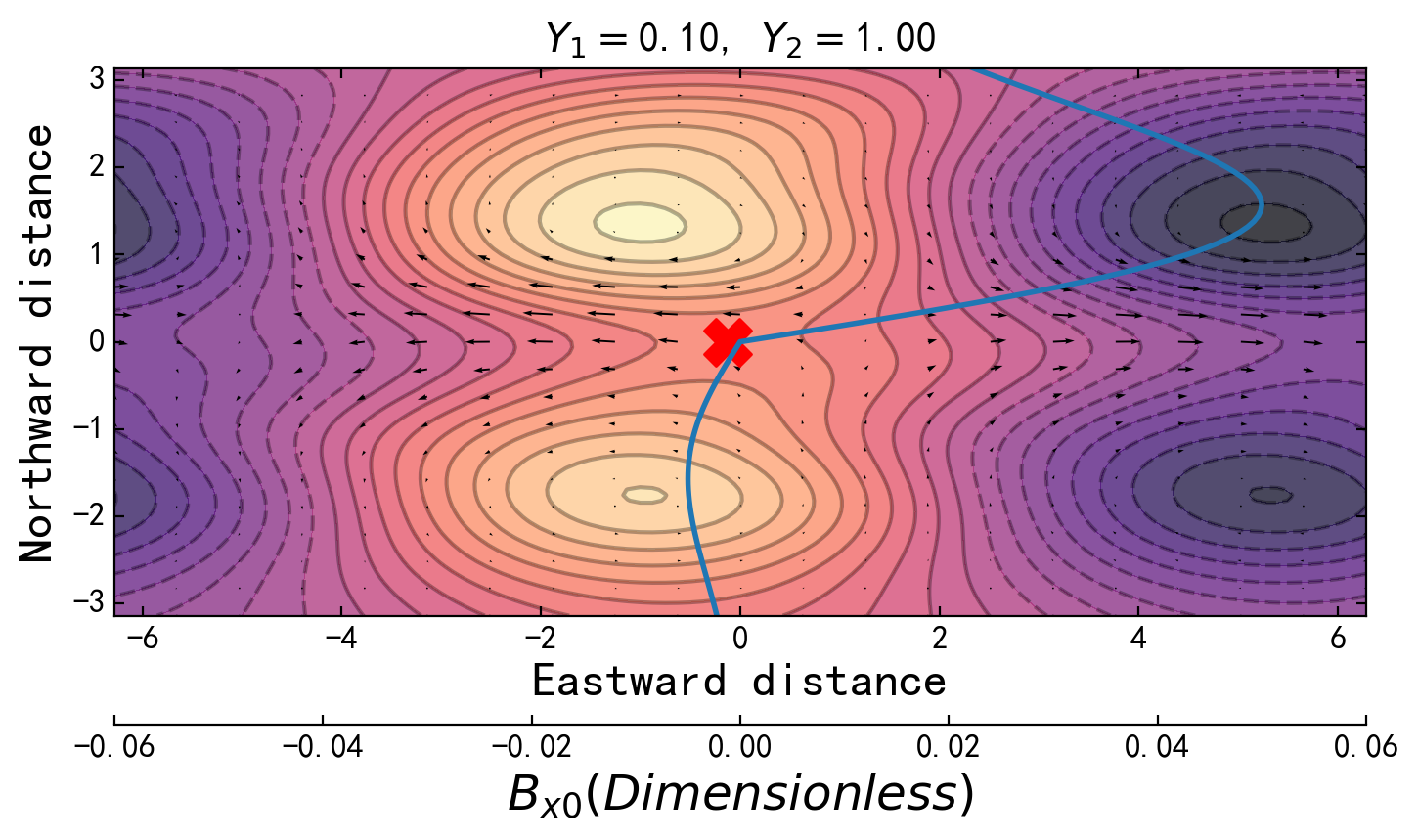}
    \end{subfigure}
    \begin{subfigure}[b]{0.33\linewidth}
        \centering
        \includegraphics[width=\linewidth]{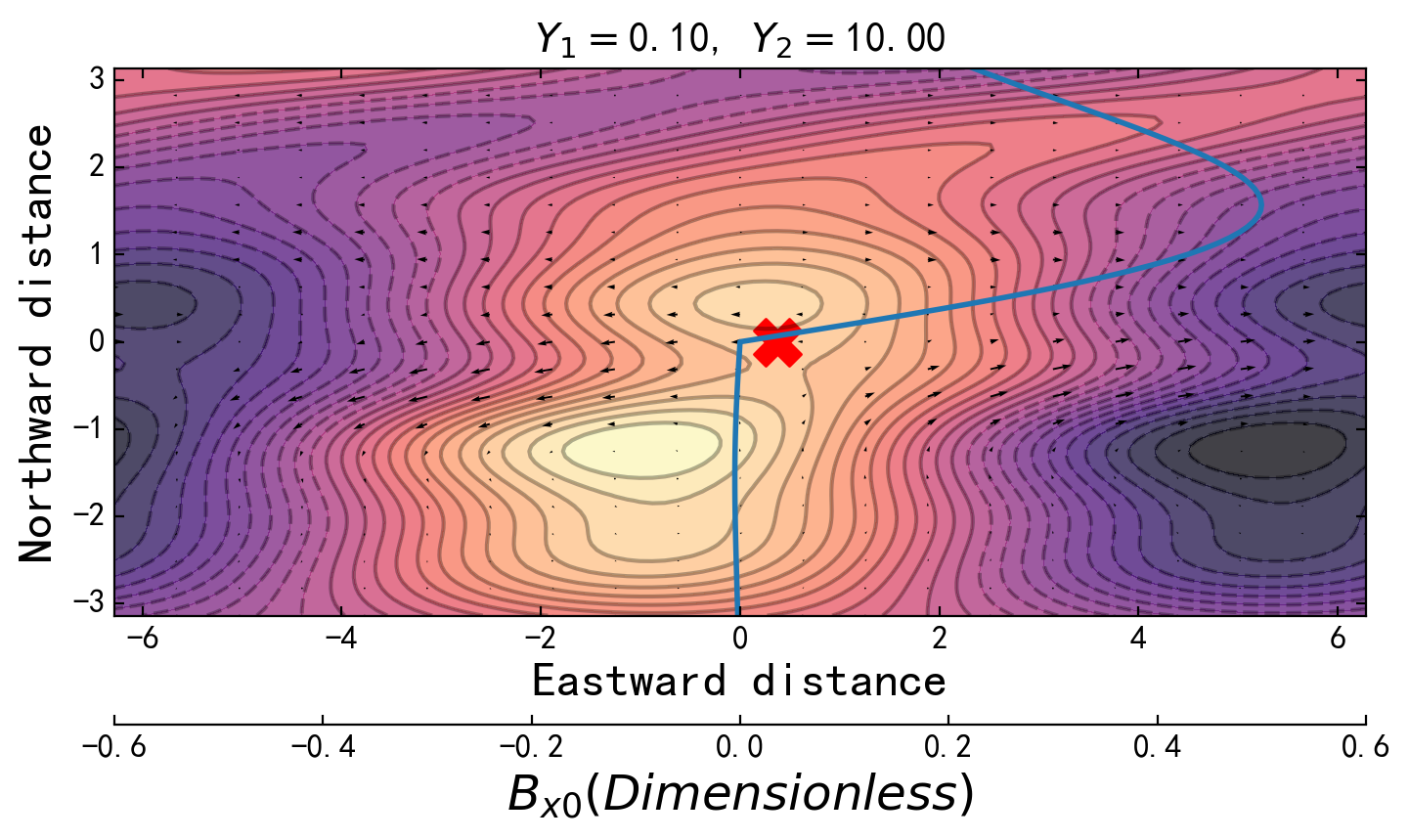}
    \end{subfigure}
    
    \begin{subfigure}[b]{0.33\linewidth}
        \centering
        \includegraphics[width=\linewidth]{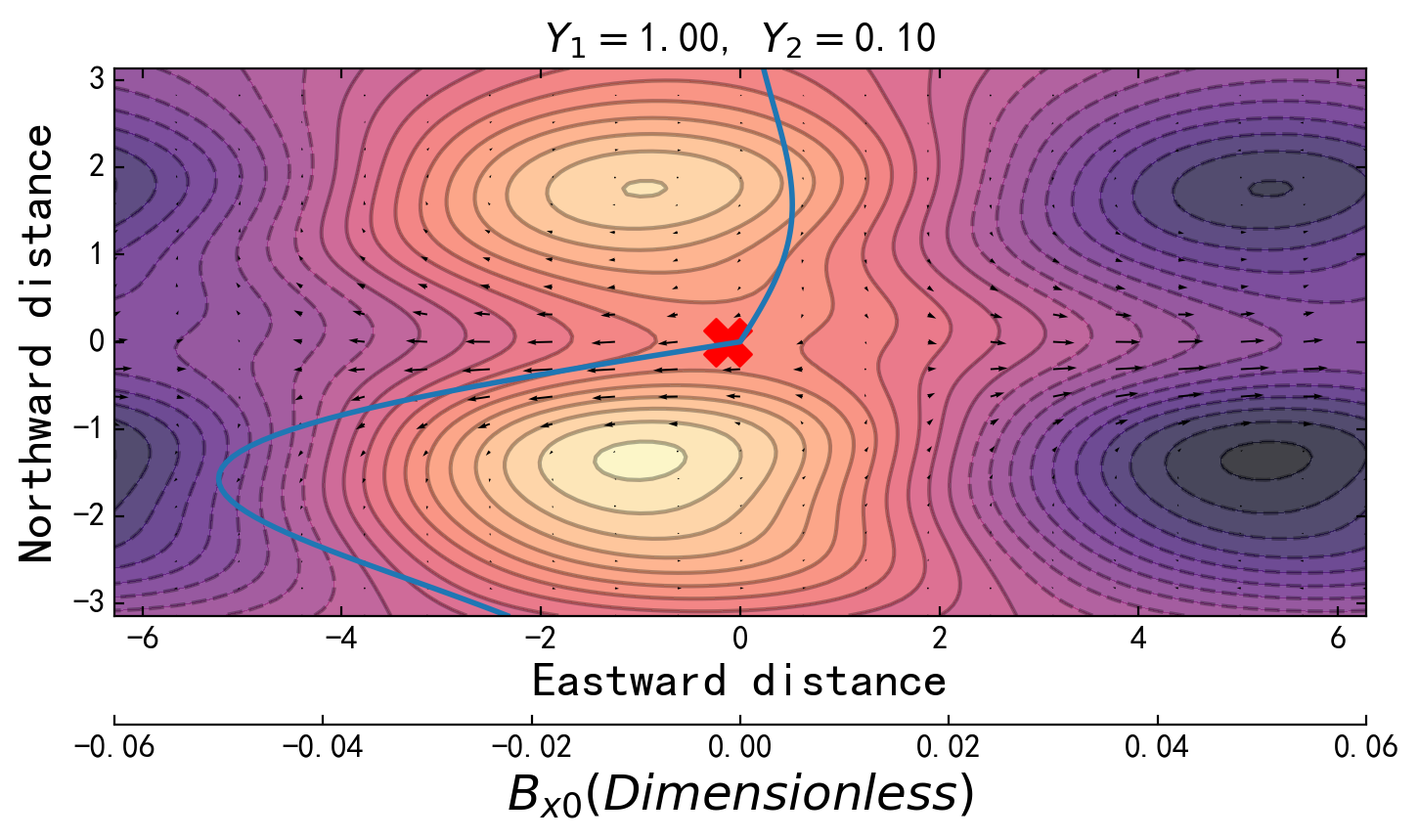}
    \end{subfigure}
    \begin{subfigure}[b]{0.33\linewidth}
        \centering
        \includegraphics[width=\linewidth]{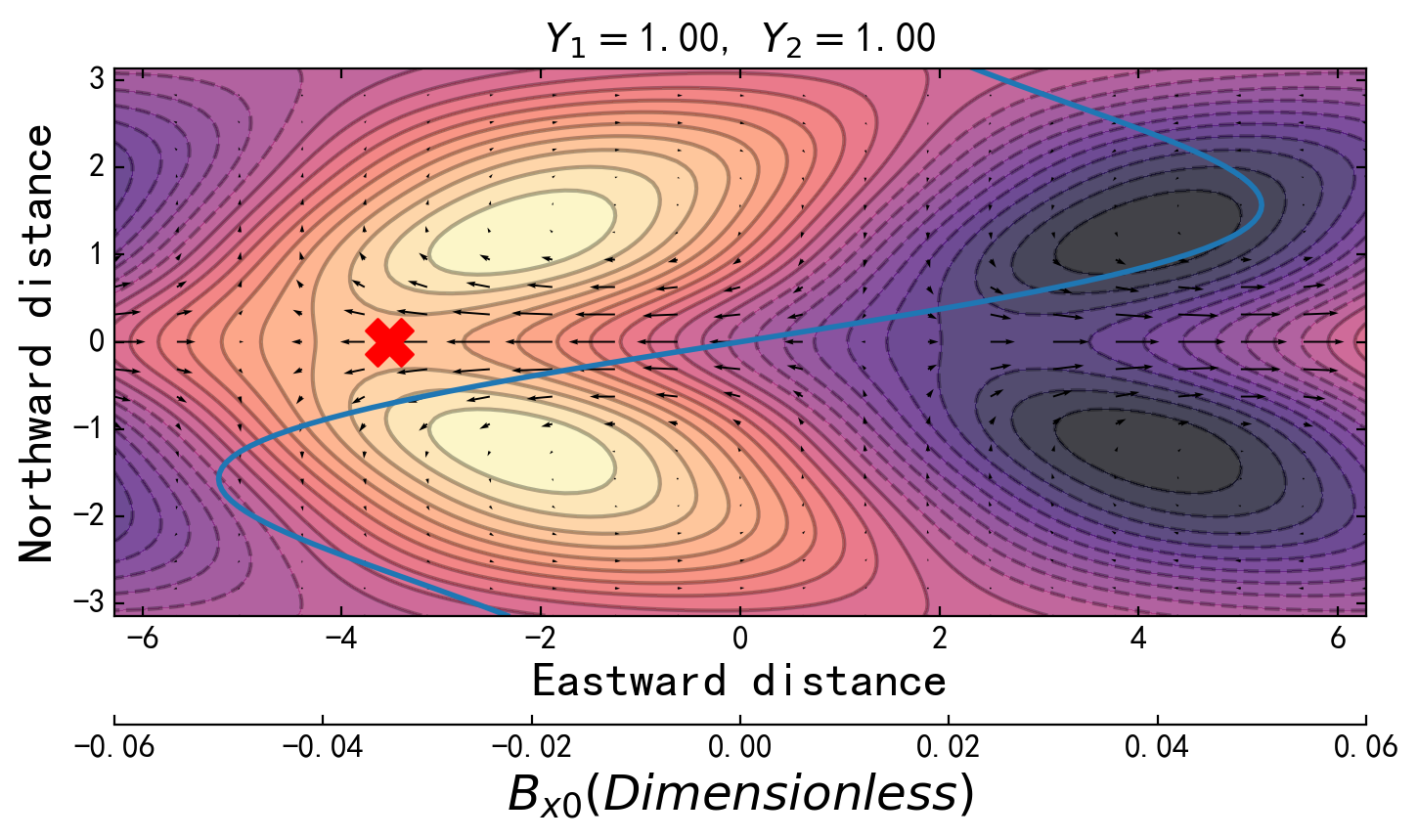}
    \end{subfigure}
    \begin{subfigure}[b]{0.33\linewidth}
        \centering
        \includegraphics[width=\linewidth]{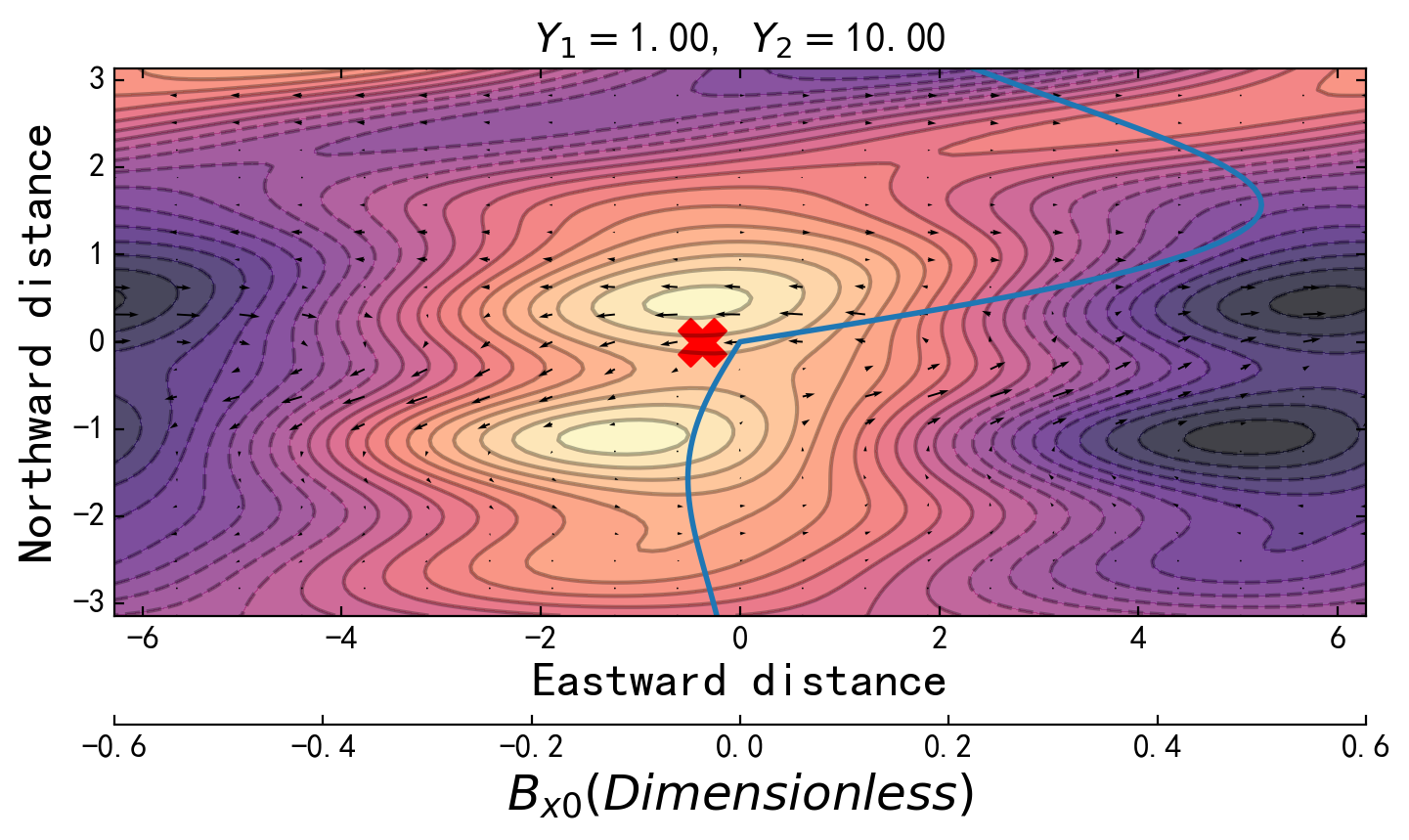}
    \end{subfigure}
    
    \begin{subfigure}[b]{0.33\linewidth}
        \centering
        \includegraphics[width=\linewidth]{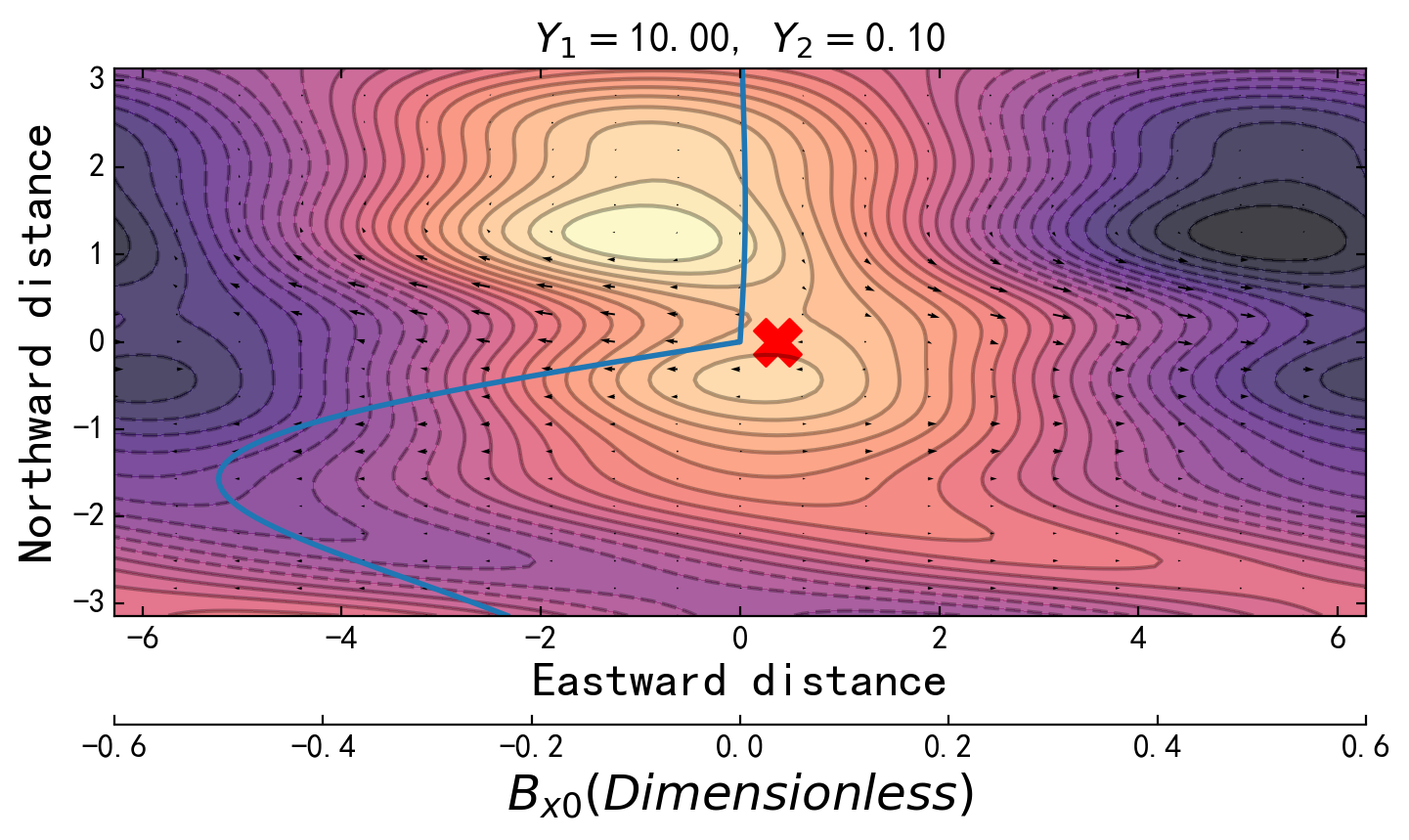}
    \end{subfigure}
    \begin{subfigure}[b]{0.33\linewidth}
        \centering
        \includegraphics[width=\linewidth]{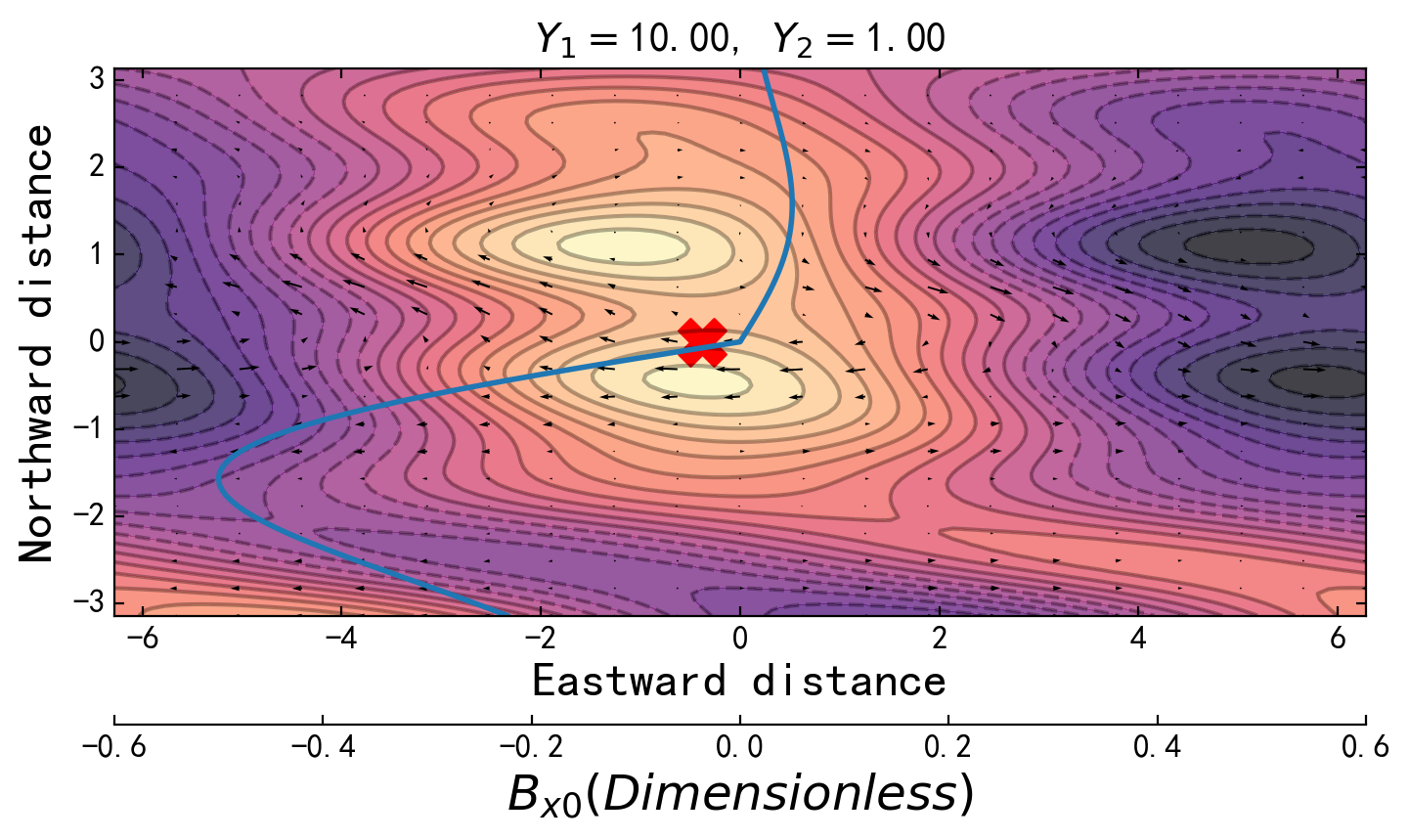}
    \end{subfigure}
    \begin{subfigure}[b]{0.33\linewidth}
        \centering
        \includegraphics[width=\linewidth]{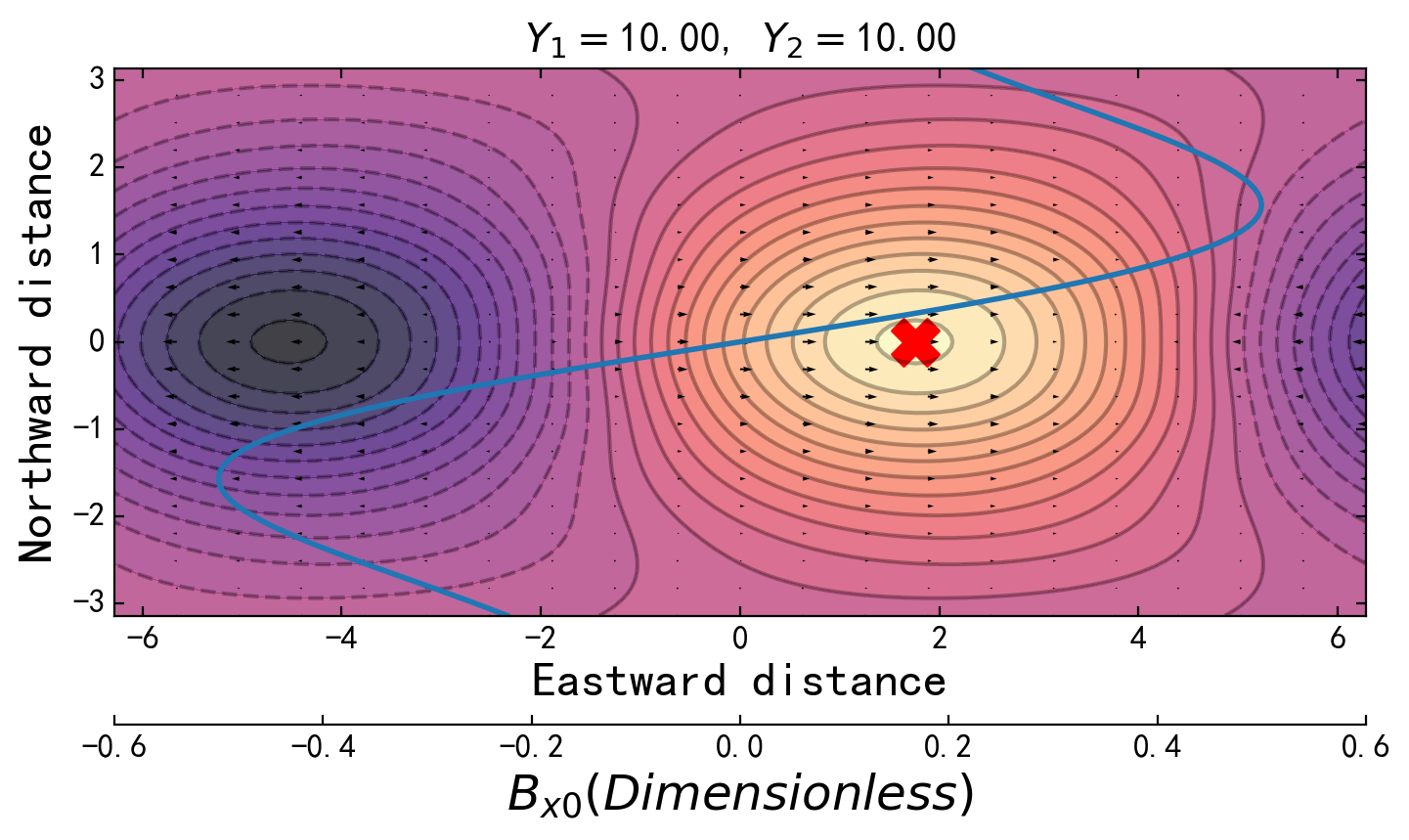}
    \end{subfigure}
    
    \caption{Height field (scale) and horizontal wind velocities (arrows) are the analytic solutions of linearized perturbed equations (equation \ref{3-6}). From top to bottom and from left to right, the results correspond to $Y_1=10, 1, 0.1$ and $Y_2=10, 1, 0.1$ respectively. The Alfvén speed $V_A=0.05$. The blue curves depict the initial magnetic field strength $B_{x0}(y)$, whose values correspond to the $B_{x0}$ axis at the bottom. The red cross marks the longitude position on the equator where $h$ reaches a maximum value.}
    \label{figure 1}
\end{figure*}

We vary the coefficients$\left(Y_1,Y_2\right)$ of equation \eqref{2-6} to obtain different north-south asymmetric magnetic fields and examine the variations in hotspot positions and temperature distributions under different conditions.  

When the magnetic field is very small (upper left corner of Figure \ref{figure 1}), its height field is similar to that in pure dynamical case presented in \cite{2011ApJ...738...71S}, equatorial hotspot moves eastward relative to the substellar point. Cyclones and anticyclones are clearly visible and exhibit height extrema that are phase-shifted westward relative to radiative equilibrium, the thickness variation along the equator becomes more moderate relative to those in mid-to-high latitude regions $\left( \left| y\right| \sim 1\text{-}3 \right)$. The pattern exhibits an overall northwest-to-southeast tilt in the northern hemisphere and a southwest-to-northeast tilt in the southern hemisphere, resembling an eastward-pointing chevron-shape centered on the equator. 
When the magnetic field has moderate strength (middle of Figure \ref{figure 1}), the equatorial hotspot moves westward, in stark contrast to the shallow-water hydrodynamic (SWHD) system (top left of Figure \ref{figure 1}). As the magnetic field is increased, the eastward-pointing chevron-shaped contours in the SWHD system undergo a phase transition and transform into westward-pointing chevron-shaped contours, the pattern tilts southwest-to-northeast in the northern hemisphere and tilts northwest-to-southeast in the southern hemisphere, which is completely opposite to the pattern in SWHD system.
When the magnetic field becomes large enough (lower right corner of Figure \ref{figure 1}), the westward-pointing chevron-shaped contours become indistinct, and the hotspot lies on the equator and close to the position of the hotspot in the purely hydrodynamic case.

The three magnetic fields mentioned above have different magnitudes but all exhibit antisymmetry at the equator. Next, we change the magnetic field strength of one hemisphere to obtain an asymmetric magnetic field configuration between the northern and southern hemispheres.
When the coefficients $Y_1=0.1$ and $Y_2=1.0$ or $Y_1=1.0$ and $Y_2=0.1$, the magnetic field strength differs by a factor of 10 between the two hemispheres, though both remain weak overall. Compared with the result in the pure dynamical case, the equatorial hotspot shifts relatively westward, moving closer to the substellar point. The hemisphere with stronger magnetic field exhibits higher temperatures on the dayside, the tilt direction of the temperature field pattern reverses, and a temperature difference emerges between the two hemispheres.
When the coefficients $Y_1=0.1$, $Y_2=10$ or $Y_1=10$, $Y_2=0.1$, the magnetic field strength difference between the hemispheres reaches a factor of 100, generating a significant asymmetric magnetic field where one hemisphere exhibits much stronger magnetic field intensity than the other. The equatorial hotspot shifts eastward but stays close to the substellar point. On the dayside, the regions of maximum temperatures in both hemispheres shift toward the equator, and regions with stronger magnetic fields lie even closer to the equator. And notably, the hemisphere of stronger magnetic field strength shows lower temperatures than the hemisphere with weaker magnetic field. 
When the coefficients $Y_1=1.0$, $Y_2=10.0$ or $Y_1=10.0$, $Y_2=1.0$, the magnetic field strength in one hemisphere becomes ten times that of the other. The equatorial hotspot shifts westward but stays close to the substellar point. The dayside temperature maxima in both hemispheres shift toward the equator, with those in the hemisphere of stronger magnetic field shifted farther eastward. This phenomenon arises from the suppressing effect of the Lorentz force on atmospheric flow, whereby asymmetric magnetic fields modify meridional heat transport.

\begin{figure}[htbp] 
    \centering    
    \begin{subfigure}{0.45\textwidth}
        \centering
        \includegraphics[height=4.5cm]{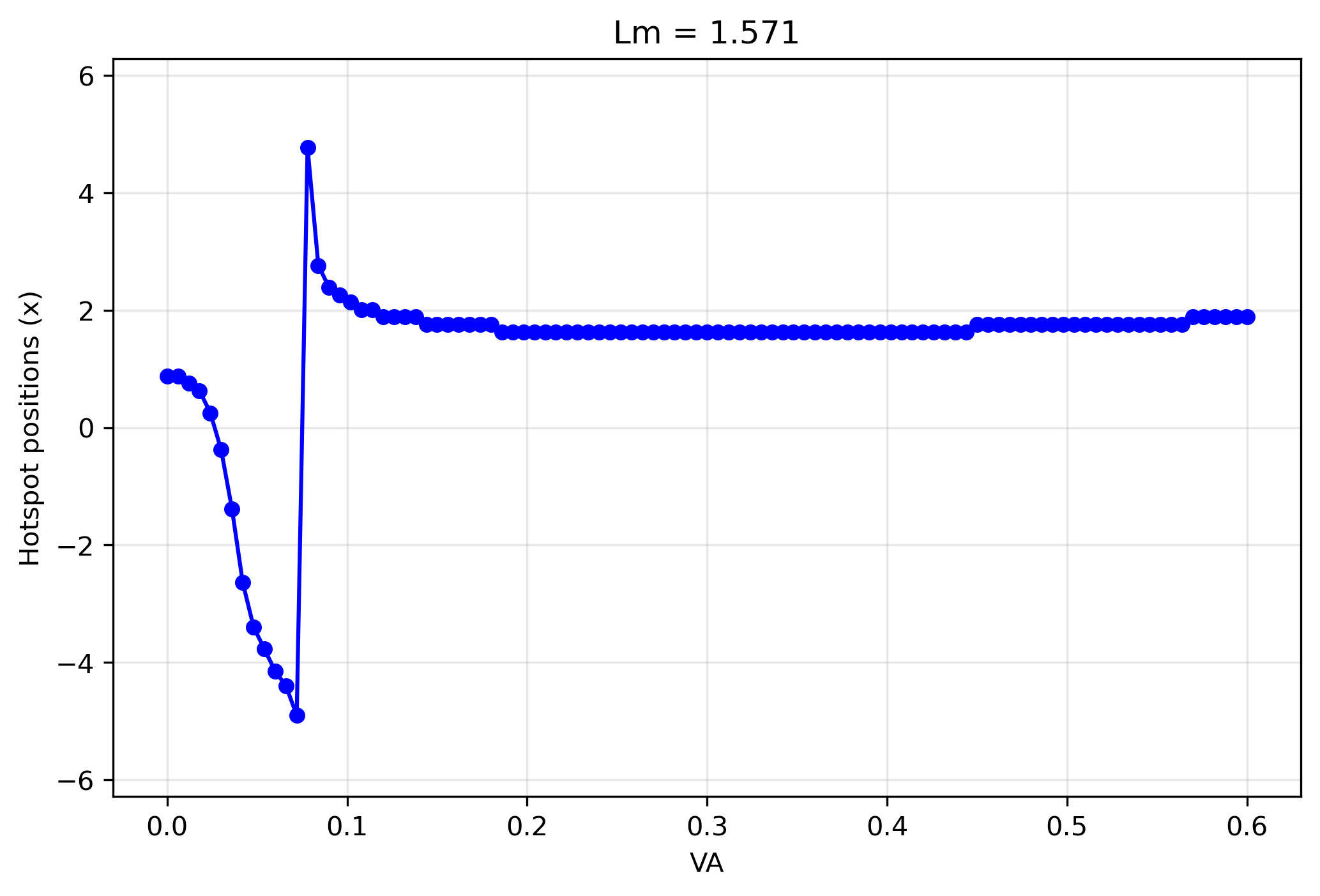} 
    \end{subfigure}
    \hfill 
    \begin{subfigure}{0.45\textwidth}
        \centering
        \includegraphics[width=\textwidth,height=4.5cm]{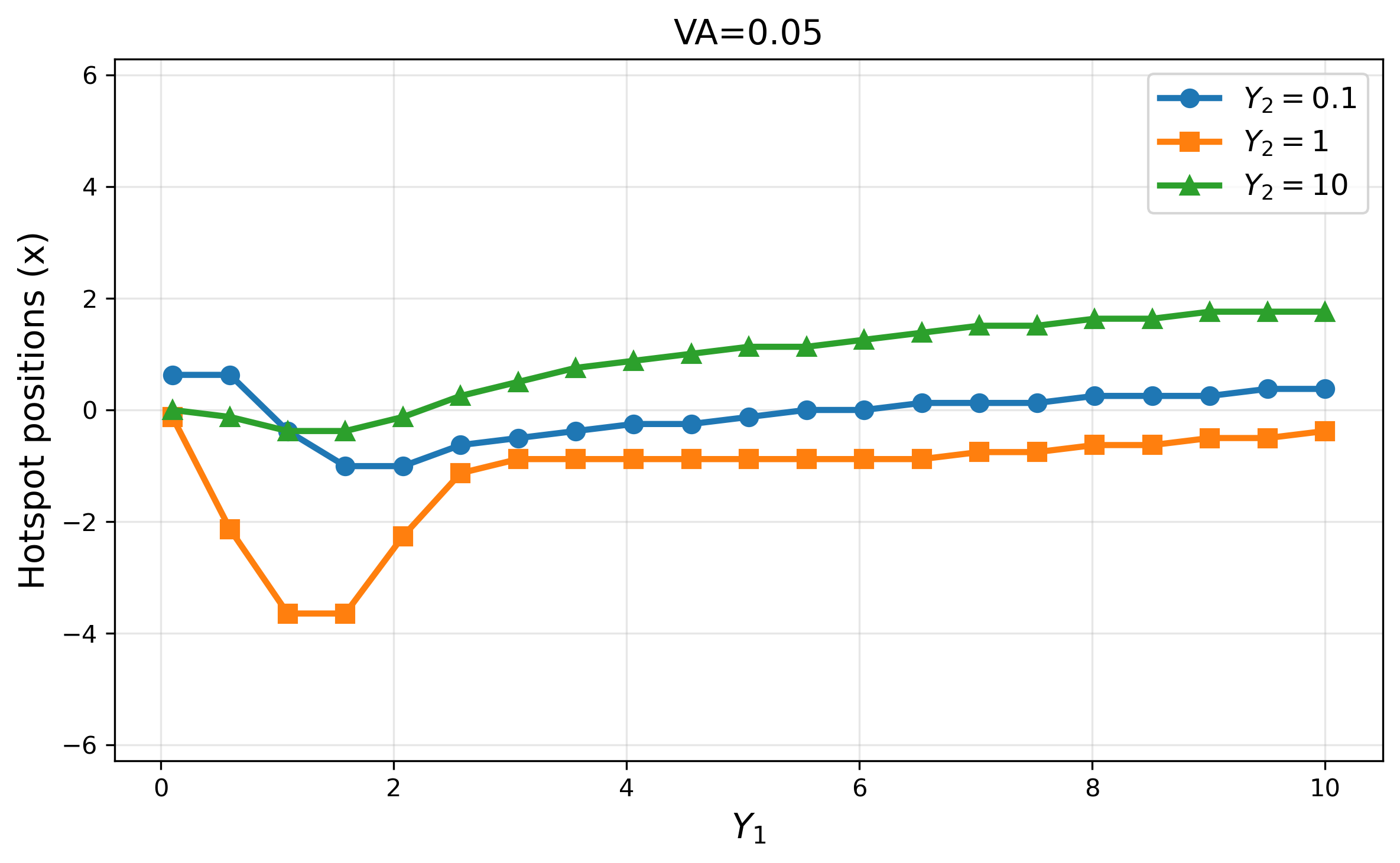}
    \end{subfigure}
    \caption{(a) Equatorial hotspot longitudinal position as a function of $V_A$. (b) Equatorial hotspot longitudinal position as a function of $Y_1$ for three values of $Y_2=0.1,1,10$, with fixed $V_A = 0.05$.}
    \label{figure 2}
\end{figure}

To clearly describe the relationship between the equatorial hotspot position and magnetic field strength, we present Figure \ref{figure 2}. First, we explore the variation of the hotspot position with $V_A$ (Figure \ref{figure 2}(a)). As $V_A$ increases, equatorial hotspots gradually move westward from the east of the substellar point, pass across the boundary, and eventually stabilize in the eastern region. The abrupt jump in the hotspot position around $V_A\approx0.08$ is induced by the periodic boundary conditions adopted in the model.
Next, we fix $V_A=0.05$ (an appropriate position west of the substellar point, chosen to ensure clear observation and avoid interference) and analyze how the equatorial hotspot position varies with $Y_1$ for $Y_2=0.1,1,10$ (Figure \ref{figure 2}(b)). As $Y_1$ increases, the equatorial hotspots first shift westward, then eastward, and eventually becomes steady. The strongest westward displacement occurs at $Y_2 =1$ and $Y_1\approx0.5-2$, indicating that a significant westward shift of the equatorial hotspot only appears when the magnetic field strength is moderate and the north–south hemispheric difference is small.

\subsection{Effects of Asymmetric Magnetic Field on Momentum Transport}
\begin{figure*}[htbp]
    \centering
    \begin{subfigure}[b]{0.33\linewidth}
        \centering
        \includegraphics[width=\linewidth]{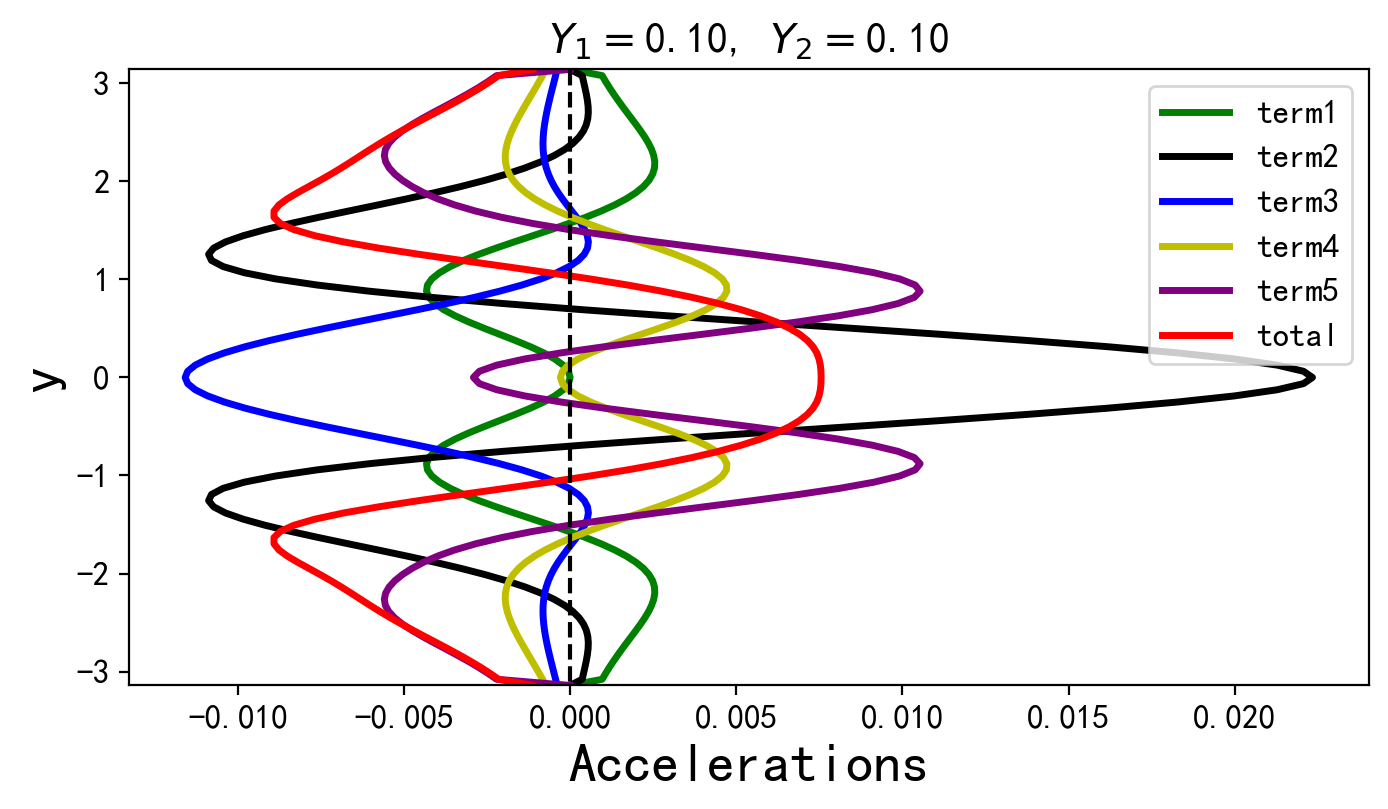}
    \end{subfigure}
    \begin{subfigure}[b]{0.33\linewidth}
        \centering
        \includegraphics[width=\linewidth]{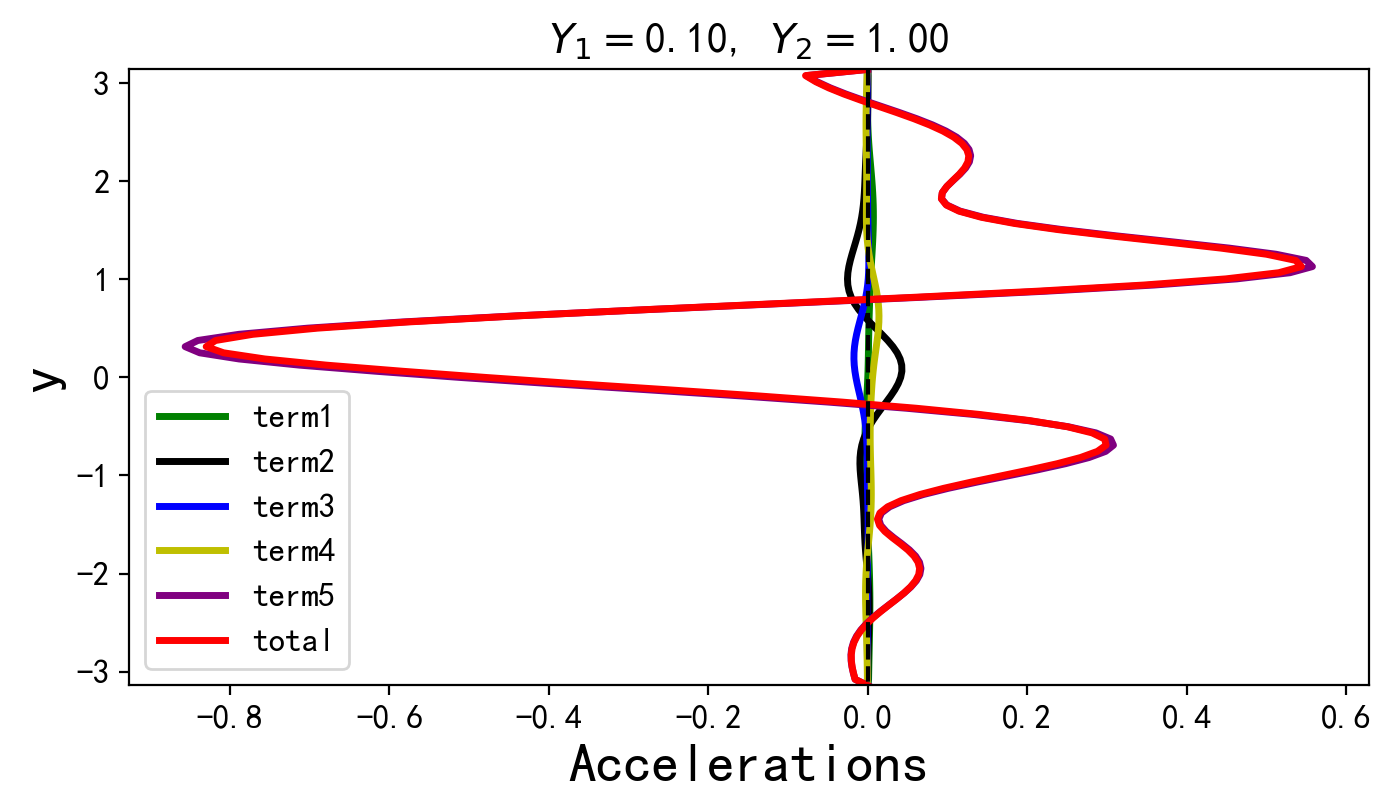}
    \end{subfigure}
    \begin{subfigure}[b]{0.33\linewidth}
        \centering
        \includegraphics[width=\linewidth]{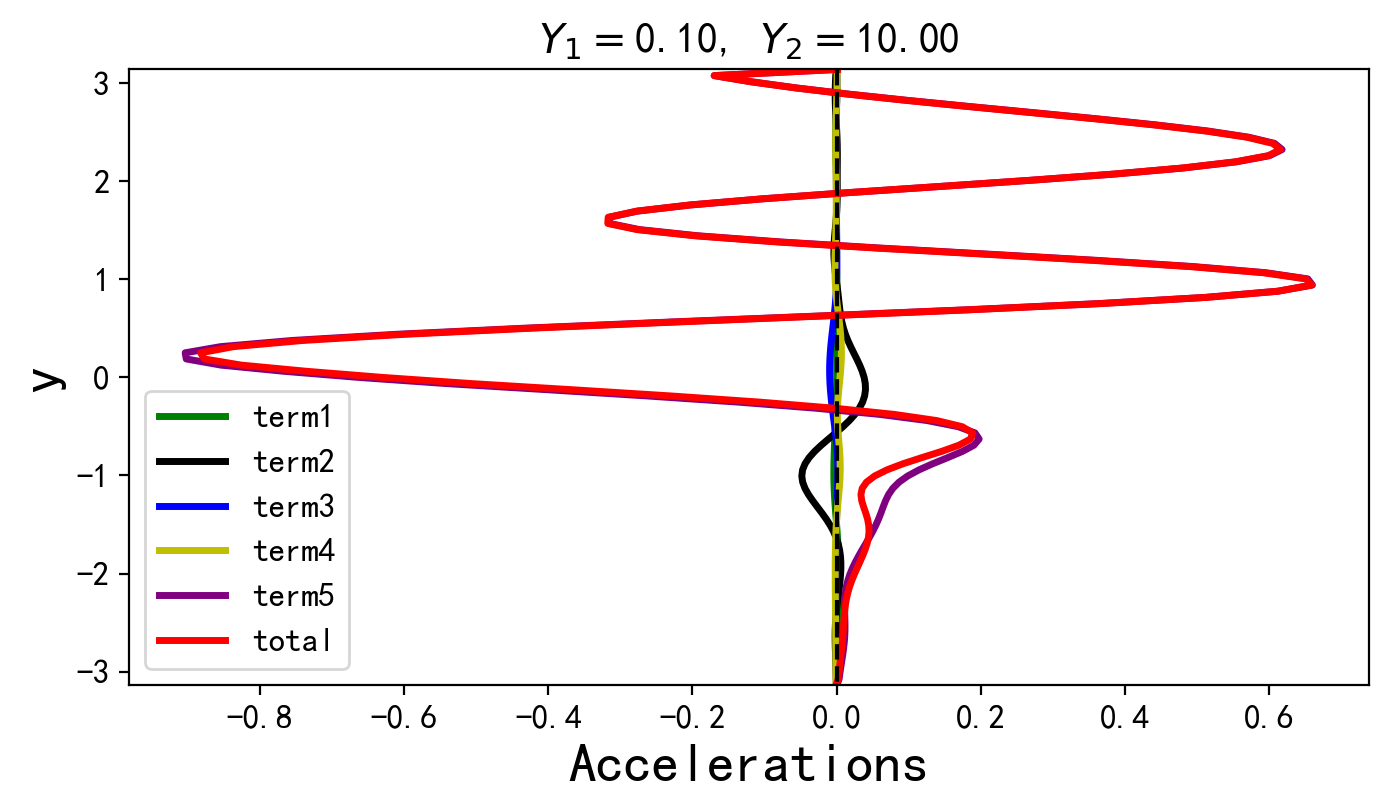}
    \end{subfigure}
    
    \begin{subfigure}[b]{0.33\linewidth}
        \centering
        \includegraphics[width=\linewidth]{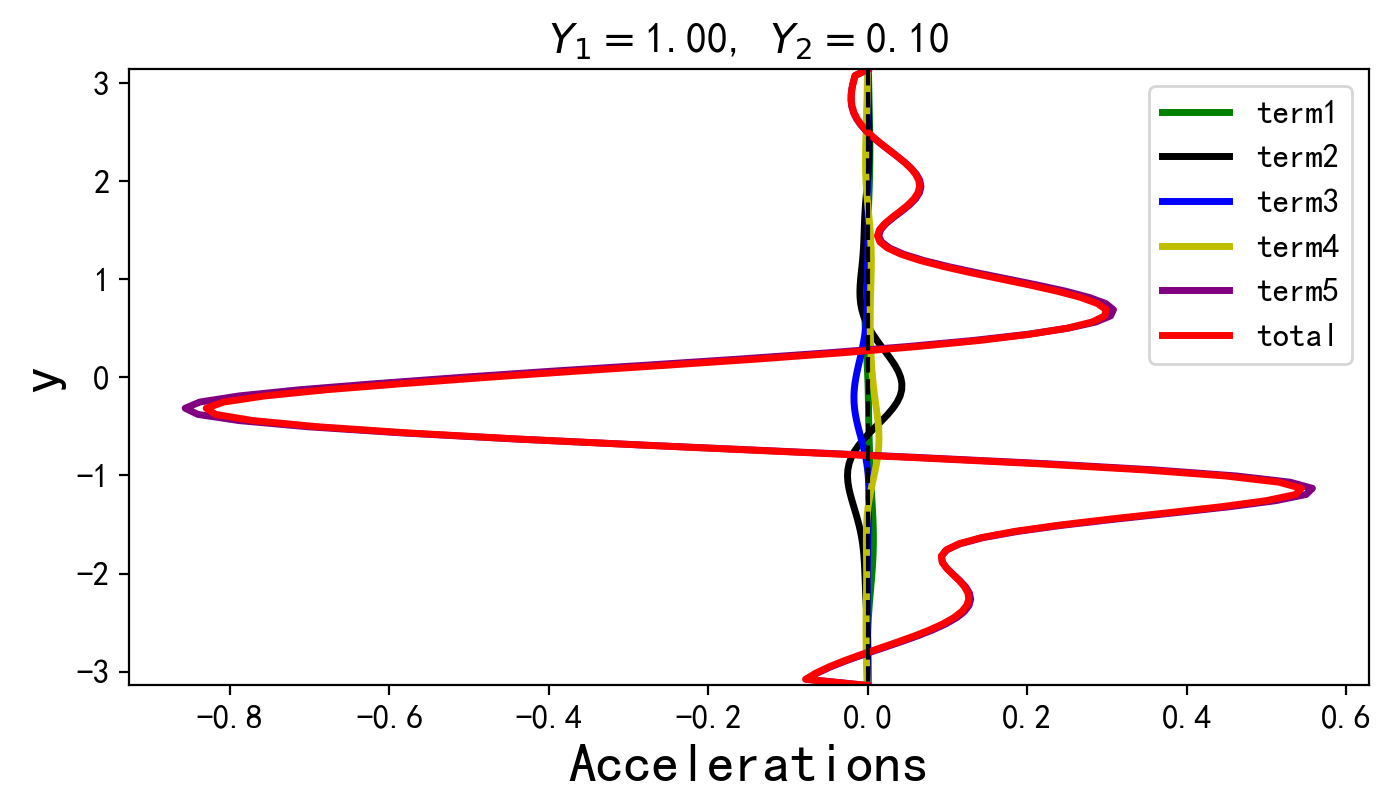}
    \end{subfigure}
    \begin{subfigure}[b]{0.33\linewidth}
        \centering
        \includegraphics[width=\linewidth]{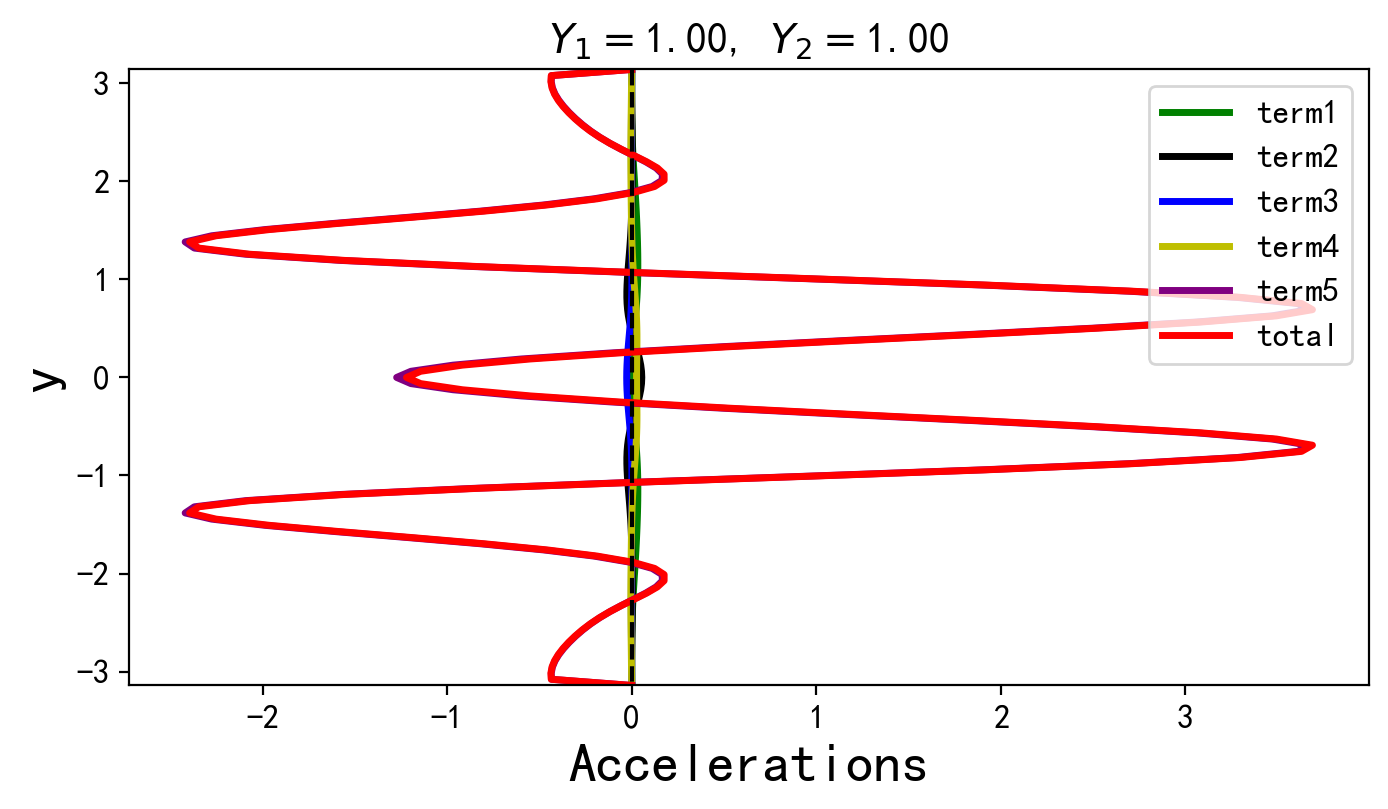}
    \end{subfigure}
    \begin{subfigure}[b]{0.33\linewidth}
        \centering
        \includegraphics[width=\linewidth]{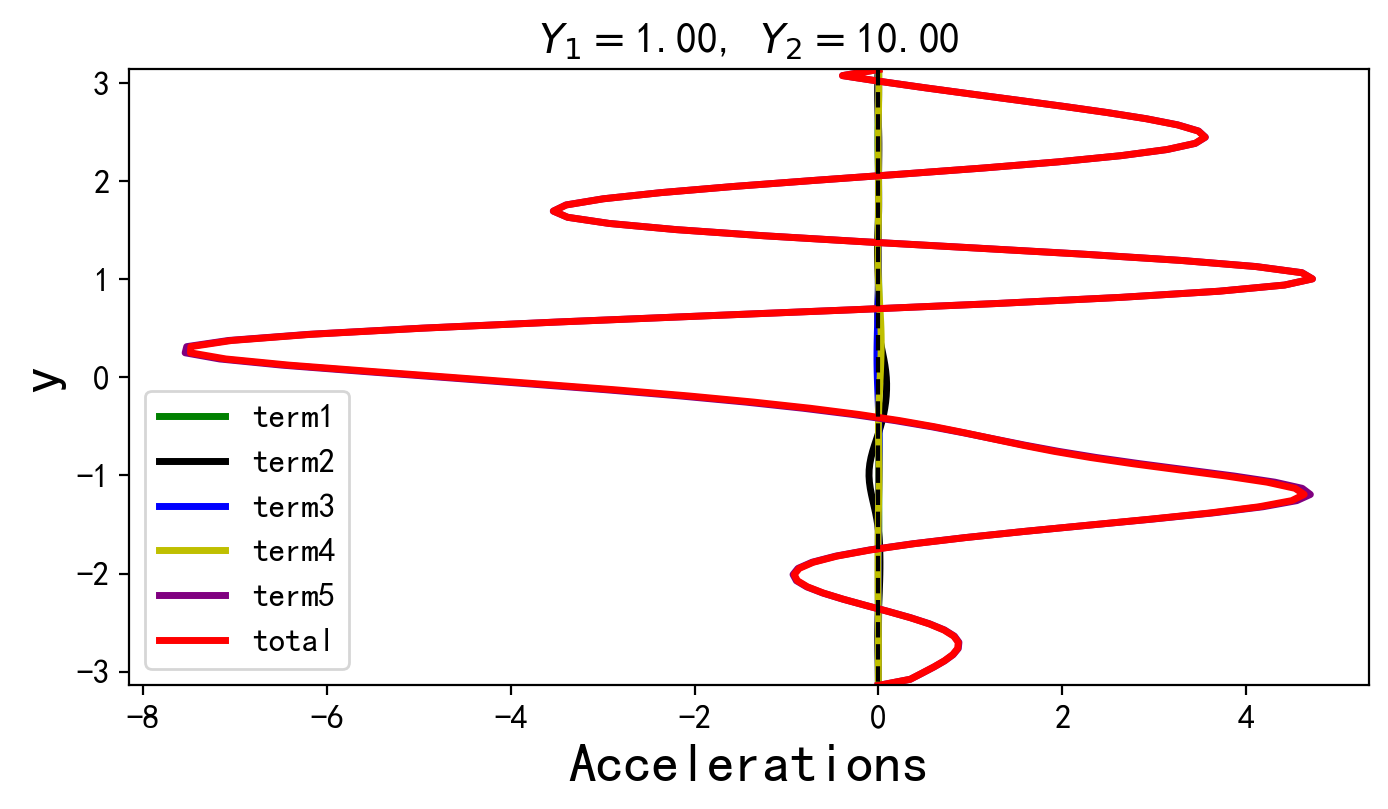}
    \end{subfigure}
    
    \begin{subfigure}[b]{0.33\linewidth}
        \centering
        \includegraphics[width=\linewidth]{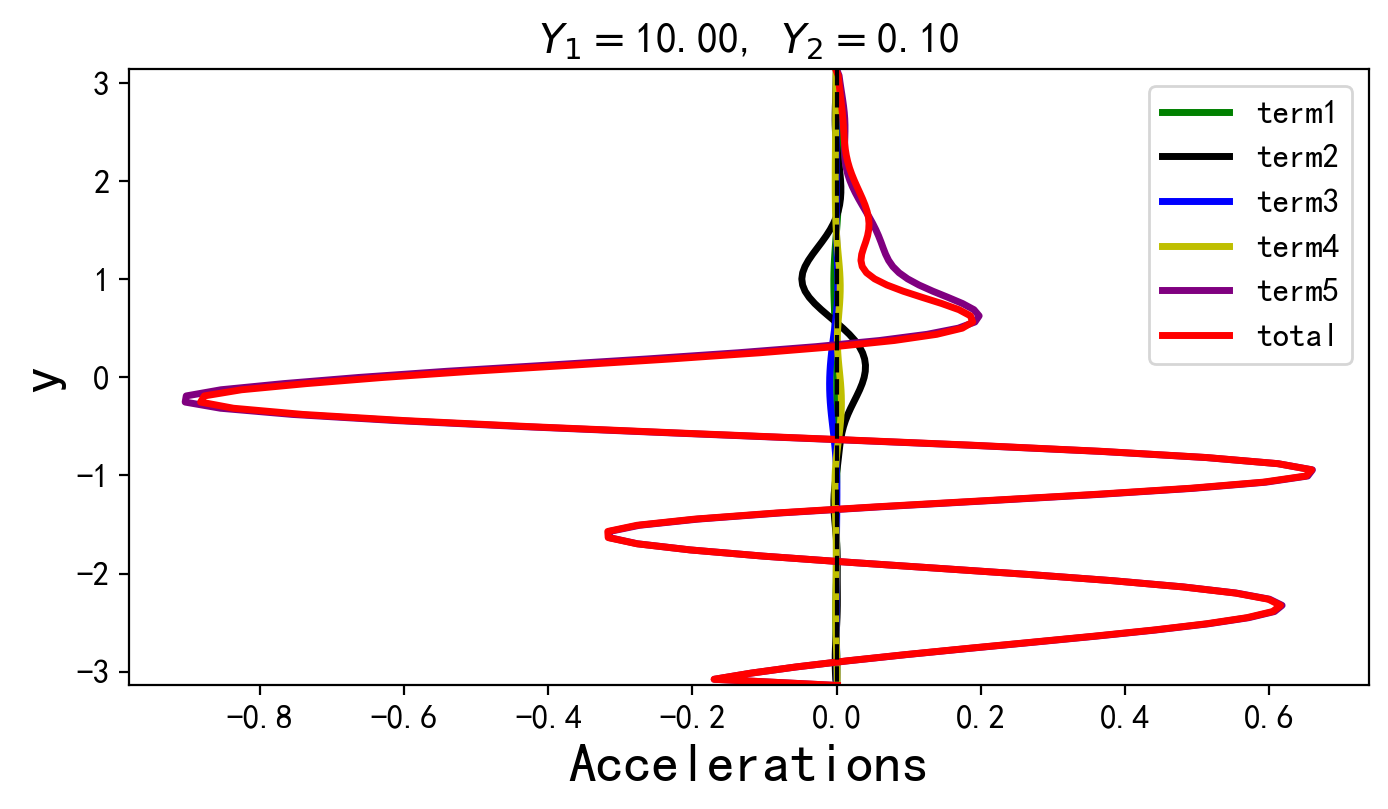}
    \end{subfigure}
    \begin{subfigure}[b]{0.33\linewidth}
        \centering
        \includegraphics[width=\linewidth]{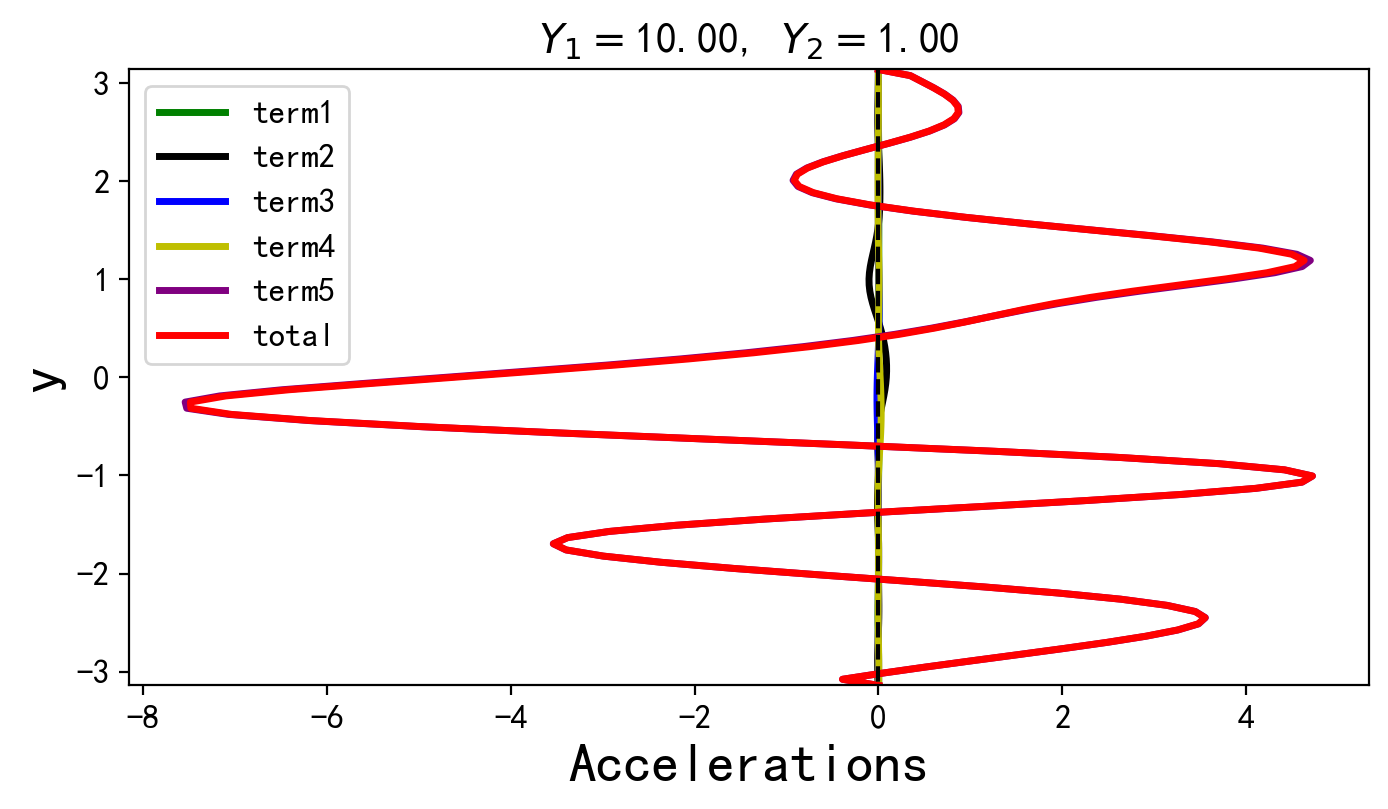}
    \end{subfigure}
    \begin{subfigure}[b]{0.33\linewidth}
        \centering
        \includegraphics[width=\linewidth]{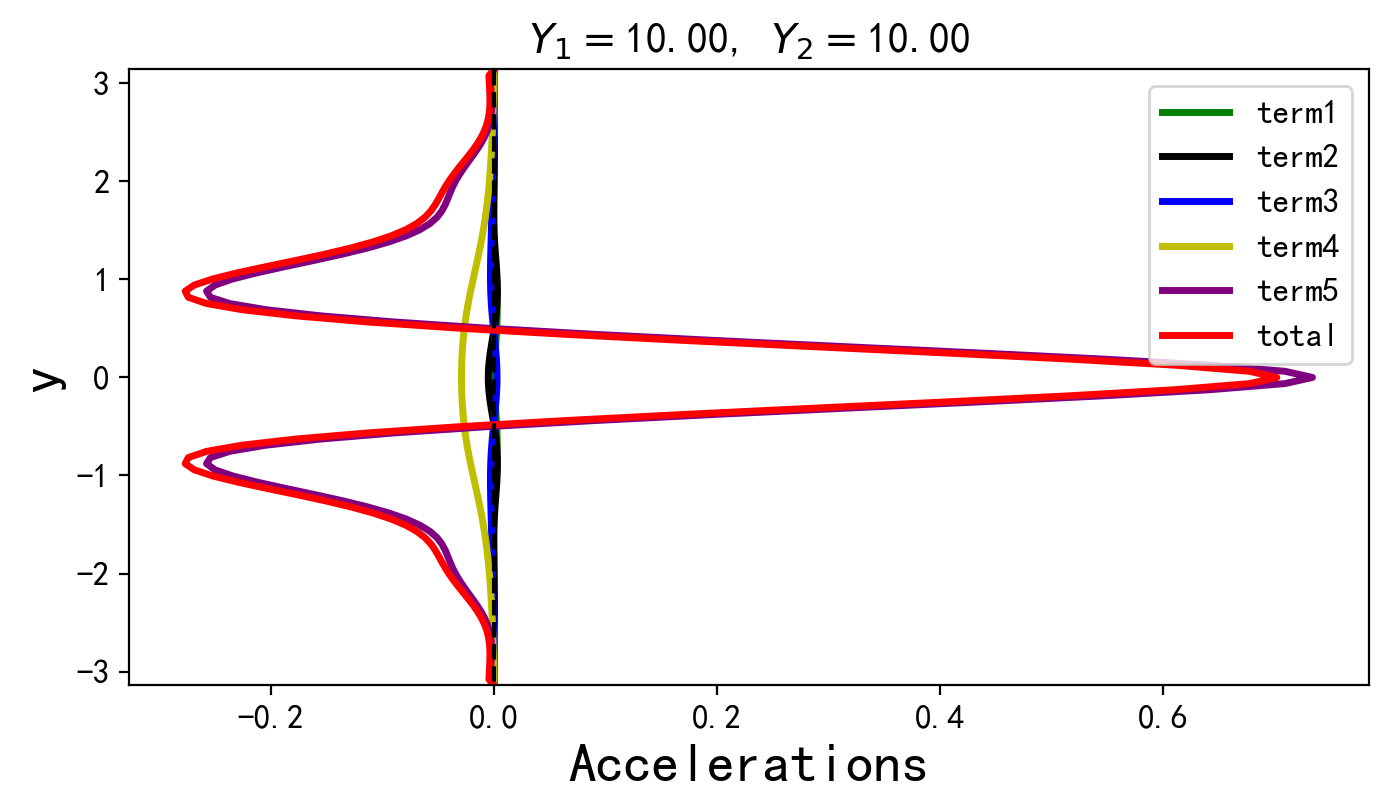}
    \end{subfigure}
    
    \caption{Zonal (east–west) accelerations of the zonal-mean flow implied by the linear solution in the figure 1. The black and dark blue curves represent accelerations induced by horizontal and vertical eddy transport (terms II and III, respectively, in Equation \ref{4-1}). The light-green and green curves show friction (term IV) and the effect of the mean meridional circulation (term I), respectively. The purple curve shows term V, which is the acceleration caused by the magnetic field. The red curve is obtained by summing all terms. }
    \label{figure 3}
\end{figure*}

To further analyze how asymmetric magnetic fields influence the atmospheric circulation of hot Jupiters, we introduce the zonal (east-west) acceleration of the zonal-mean flow to investigate the effects of magnetic fields on momentum transport. By decomposing the variables in the momentum equation (equation \eqref{2-1}) into zonal mean components (denoted with an overbar, e.g., $\overline{u}$) and perturbations (denoted with a prime symbol, e.g., $u'$) and taking the zonal (x-direction) average, we can obtain the zonal acceleration equation for the zonal mean flow, which is the magneto-hydrodynamic shallow-water version of the Transformed Eulerian Mean (TEM) momentum equation \citep{1987mad..book.....A}:
\begin{equation}
\begin{aligned}
\frac{\partial \overline{u}}{\partial t} = 
\underbrace{ \overline{V_y}^{*}\Bigg[f-\frac{\partial \overline{u}}{\partial y}\Bigg]}_I -
\underbrace{ \frac{1}{\overline{h}}\frac{\partial}{\partial y}[\overline{(hv)'u'}]}_{II} +
\underbrace{\Bigg[\frac{1}{\overline{h}}\overline{u'Q'}+\overline{R_u}^{*}\Bigg]}_{III} \\
-\underbrace{\frac{\overline{V_x}^{*}}{\tau_{\rm drag}}}_{IV}+
\underbrace{\frac{1}{\overline{h}}\overline{\Bigg[\overline{\left(hB_y\right)}\frac{\partial \overline{b_x}}{\partial y}+\frac{1}{\overline{h}}\frac{\partial [\overline{(hB_y)'b_x'}]}{\partial y}\Bigg]} }_{V}-
\frac{1}{\overline{h}}\frac{\partial(\overline{h'u'})}{\partial t}.
\end{aligned}
\tag{4-1}
\label{4-1}
\end{equation}
$\overline{M}^{*} \equiv \overline{hM}/\overline{h}$ represents the thickness-weighted zonal mean of any quantity $M$. The left-hand side of the equation corresponds to the acceleration of the zonal mean velocity, while the right-hand side comprises terms I-V which correspond to accelerations caused by: momentum advection due to the mean meridional circulation; convergence of the meridional flux of the zonal eddy momentum; radial (vertical) eddy momentum transport; drag force; magnetic field effects. The final term denotes the time rate of change of eddy momentum. Terms I-II are primarily associated with meridional momentum transport, while term V reflects the magnetic field effects. By plotting terms I-V individually as functions of $y$ and summing them to obtain the total term, we can compare the contributions of each term to the acceleration of the mean flow and analyze their combined effect (e.g., as illustrated in Figure \ref{figure 3}).

When the magnetic field strength is very small (top left of Figure \ref{figure 3}), a net eastward acceleration is observed at the equator. Term II, the horizontal convergence of eddy momentum flux, generates strong eastward acceleration at the equator and westward acceleration at mid-latitudes (black curve). Term III, associated with vertical eddy momentum transport (blue curve), produces a strong westward acceleration at the equator, implying downward transport of eddy momentum there. The mean-meridional circulation (Term I, green curve) and mass-weighted friction (Term IV, light green curve) are small near the equator and largely cancel each other. Term V contributes a minor westward acceleration at the equator. Thus, Term II plays the main role, resulting in net eastward acceleration at the equator and westward acceleration at mid-latitudes (red curve).
When the magnetic field strength increases (middle of Figure \ref{figure 3}), the accelerations produced by all terms grow significantly. Term V (the magnetic field term) exceeds the sum of all other terms and generates a westward acceleration at the equator. 
As the magnetic field strength becomes extremely high, the acceleration decreases and generates strong eastward acceleration at the equator and westward acceleration at mid-latitudes, thus allowing the height field to return to radiation balance.

After introducing an asymmetric magnetic field, the accelerations in the northern and southern hemispheres become distinctly asymmetric. When the magnetic field is sufficiently strong in one hemisphere, the magnetic term becomes the dominant mechanism governing momentum transport.
When $Y_1=0.1$, $Y_2=1.0$ and $Y_1=1.0$, $Y_2=0.1$, we compare the two figures and find that the distribution of acceleration is completely opposite in the northern and southern hemispheres, consistent with the distribution of magnetic field. The total acceleration is westward at the equator. The hemisphere with a relatively stronger magnetic field exhibits a larger acceleration. Term II causes westward acceleration at mid-latitudes in the hemisphere with stronger magnetic field, indicating that westward momentum is transported meridionally to the mid-latitudes, corresponding to the westward shift and intensification of cyclones in that hemisphere.  
When the magnetic field strength difference between the northern and southern hemispheres becomes 100 times (middle right of Figure \ref{figure 3}), the acceleration distribution shows a significant change. The acceleration in the hemisphere with the stronger magnetic field overall appears to be larger and varies significantly with latitude. The westward acceleration generated by Term II in the mid-latitudes of the hemisphere with the weaker magnetic field implies that westward momentum is transported meridionally into mid-latitudes regions, leading to the westward shift and intensification of cyclones in that hemisphere.
When $Y_1=1$, $Y_2=10$ and $Y_1=10$, $Y_2=1$, the acceleration increases significantly and the maximum acceleration occurs in the hemisphere with the stronger magnetic field near the equator. 

Even if only one hemisphere’s magnetic field is strengthened, term V (magnetic field) dominates the zonal acceleration. The magnitude of the total acceleration approaches that of term V as the magnetic field strengthens. At the equator, the net acceleration direction matches the motion direction of the equatorial hotspot. In the hemisphere with a stronger magnetic field, the zonal-mean acceleration of the zonal-mean flow is also larger.

\section{Discussion}
We investigate the influence of asymmetric magnetic fields on atmospheric circulation and hotspot dynamics in hot Jupiter atmospheres using a shallow-water magnetohydrodynamic (SWMHD) model in this study. Both the strength and symmetry of planetary magnetic fields significantly alter hotspot offsets and temperature distributions. 

When the magnetic field is equatorially antisymmetric but symmetric across hemispheres, increasing magnetic field strength causes the equatorial hotspot to transition from eastward to westward movement. Stronger magnetic fields generate westward drag that shifts the hotspot. This aligns with previous SWMHD simulations but contrasts with hydrodynamic models predicting persistent eastward jets.

After introducing a north-south asymmetric magnetic field, a significant temperature difference emerges between the two hemispheres. Under the influence of magnetic drag, the regions of maximum temperature on the dayside of both hemispheres shift toward the equator, and the hemisphere with a stronger magnetic field lies closer to the equatorial region. Keeping the magnetic field fixed in one hemisphere, as the field strength in the other hemisphere increases, the equatorial hotspot first shifts westward, then moves eastward, and eventually stabilizes. The pronounced westward shift of the equatorial hotspot appears only under moderate magnetic field strength and weak north–south hemispheric asymmetry. Asymmetric magnetic fields may generate more complex mechanisms that influence atmospheric circulation.

The magnetic field term (Term V in the momentum budget) dominates the equatorial acceleration in the zonal acceleration curve. A north-south asymmetric magnetic field leads to complex variations in zonal acceleration, redistributes momentum between the northern and southern hemispheres, and produces an asymmetric north-south temperature field.

These results demonstrate the critical role of magnetic field strength and geometric asymmetry in explaining the observed westward hotspot shifts, providing a theoretical foundation for future three-dimensional magnetohydrodynamic models and spectroscopic observations of exoplanetary atmospheres.

Although this study focuses on the modulation of atmospheric circulation by the horizontal magnetic field and neglects the effects of vertical magnetic field, vertical gradients of the magnetic field may have significant impacts on the atmospheric dynamics of hot Jupiters. Magnetic fields play an essential role in altering the wind profiles and shaping magnetic induction in hot Jupiter atmospheres \citep{2014ApJ...782L...4R,2014ApJ...794..132R}. Through magnetic tension, vertical field gradients may modulate vertical wind shear and vertical momentum transport, thereby affecting the vertical temperature structure of the atmosphere. It will be important for future work to develop three-dimensional magnetohydrodynamic models that fully incorporate vertical magnetic field gradients to further investigate their influence on atmospheric circulation.

\section*{Acknowledgments}
This work is supported by the National SKA Program of China (grant No. 2022SKA0120101), the National Natural Science Foundation of China (NSFC, No.12288102, 12403071), the science research grants from the China Manned Space Project (No. CMS-CSST-2021-B09, CMS-CSST-2021-B12, and CMS-CSST-2021-A10), International Centre of Supernovae, Yunnan Key Laboratory (No.202302AN360001), and the grants from The Macau Science and Technology Development Fund, and opening fund of State Key Laboratory of Lunar and Planetary Sciences (Macau University of Science and Technology) (Macau FDCT Grant No. SKL-LPS(MUST)-2021-2023). C.Y. has been supported by the National Natural Science Foundation of China (grants 12373071).

\appendix
\section{SWMHD model}
For clarity, we use Cartesian coordinates and the Navier-Stokes equation (without molecular viscosity) defined only in the horizontal. 
\begin{equation}
  \frac{d\textbf{V}}{dt} + g\nabla h = -2 \bm{\Omega}\times \textbf{V}  + \textbf{F}_{\rm others} .
  \tag{A1}\label{A1}
\end{equation}
We use $\beta$-plane approximation to rewrite the Coriolis force as 
\begin{equation}
    -2\bm{\Omega}\times \textbf{V} = -\textbf{f}\times\textbf{V}=\beta y \left( \textbf{k}\times \textbf{V}\right).
    \tag{A2}\label{A2}
\end{equation}
Next, we add other physical effects to equation \eqref{A1}:
\begin{equation}
    \textbf{F}_{\rm others}=\textbf{R}-\frac{\textbf{V}}{\tau_{\rm drag}}+\frac{(\nabla\times\textbf{B})\times\textbf{B}}{4\pi\rho }=\textbf{R}-\frac{\textbf{V}}{\tau_{\rm drag}}+\frac{\left(\textbf{B}\cdot\nabla\right)\textbf{B}}{4\pi\rho}+\frac{1}{8\pi\rho }\nabla\left(\textbf{B}\cdot\textbf{B}\right).
    \tag{A3}\label{A3}
\end{equation}
The magnetic pressure $P_B=B^2/8\pi$ is typically much smaller than the gas pressure in the infrared photosphere, thus the magnetic pressure gradient term can be neglected. Introducing the Alfvén speed 
\begin{equation}
    {\bf V}_{\rm A} = \frac{\textbf{B}}{\sqrt{4\pi\rho}},
    \tag{A4}\label{A4}
\end{equation}
so the magnetic tension can be expressed as
\begin{equation}
    \frac{\left(\textbf{B}\cdot\nabla\right)\textbf{B}}{4\pi\rho }=\left(\textbf{V}_{\rm A}\cdot\nabla\right) \textbf{V}_{\rm A}.
    \tag{A5}\label{A5}
\end{equation}
The  Alfvén speed is directly related to the strength of the magnetic field. We replace $V_A$ with $B$ in units of velocity to directly express the magnetic tension. Then the momentum equation is written as:  
\begin{equation}
\frac{d\textbf{V}}{dt} + g\nabla h + \beta y (\textbf{k}\times\textbf{V})  = \textbf{R}+ (\textbf{B}\cdot\nabla)\textbf{B}-\frac{\textbf{V}}{\tau_{\rm drag}}.
\tag{A6}\label{A6}
\end{equation}
Note that $B_x=\frac{1}{h}\frac{\partial A}{\partial y}$ and $B_y=-\frac{1}{h}\frac{\partial A}{\partial x} $.The $x$- and $y$- components of equation\eqref{A6} are:
\begin{equation}
\begin{aligned}
\frac{\partial V_x}{\partial t}+V_x\frac{\partial V_x}{\partial x} + { V_y \frac{\partial V_x}{\partial y}}+g\frac{\partial h}{\partial x}-\beta y V_y-\left(\frac{1}{h}\frac{\partial A}{\partial y}\frac{\partial}{\partial x}-\frac{1}{h}\frac{\partial A}{\partial x}\frac{\partial}{\partial y}\right)\frac{1}{h}\frac{\partial A}{\partial y}+\frac{QV_x}{h}+\frac{V_x}{\tau_{\rm drag}}&= 0,\\
\frac{\partial V_y}{\partial t}+{ V_x \frac{\partial V_y}{\partial x}} + V_y\frac{\partial V_y}{\partial y}+g\frac{\partial h}{\partial y}+\beta yV_x-\left(\frac{1}{h}\frac{\partial A}{\partial y}\frac{\partial}{\partial x}-\frac{1}{h}\frac{\partial A}{\partial x}\frac{\partial}{\partial y}\right)\left(-\frac{1}{h}\frac{\partial A}{\partial x}\right)+\frac{QV_y}{h}+\frac{V_y}{\tau_{\rm drag}}&= 0.
\end{aligned}
\tag{A7}\label{A7}
\end{equation}

Consider small perturbations in equations \eqref{3-1} and set $V_{x0}=V_{y0}=0$ to get the following linearized equations:
\begin{equation}
\begin{aligned}
\frac{\partial u}{\partial t}+g\frac{\partial \delta h}{\partial x}-\beta yv-\Bigg[-\frac{1}{H^3}\left(\frac{\partial A_0}{\partial y}\right)^2\frac{\partial \delta h}{\partial x}+\frac{1}{H^2}\frac{\partial A_0}{\partial y}\frac{\partial^2 a}{\partial x\partial y}-\frac{1}{H^2}\frac{\partial^2A_0}{\partial y^2}\frac{\partial a}{\partial x}\Bigg]+\frac{u}{\tau_{\rm drag}}&=0,\\
\frac{\partial v}{\partial t}+g\frac{\partial \delta h}{\partial y}+\beta y u+\frac{1}{H^2}\frac{\partial A_0}{\partial y}\frac{\partial^2a}{\partial x^2}+\frac{v}{\tau_{\rm drag}}&=0.
\end{aligned}
\tag{A8}\label{A8}
\end{equation}
$R$ is quadratic in the forcing amplitude, which involves the velocity and forcing amplitude, and consequently, it does not appear in the linearized equations \citep{2011ApJ...738...71S}.

The continuity equation is defined only in the horizontal directions. Its linearized form:
\begin{equation}
\frac{\partial \delta h}{\partial t} + H\left(\frac{\partial u}{\partial x}+\frac{\partial v}{\partial y}\right)= \frac{h_{\rm eq}(x,y)-\delta h-H}{\tau_{\rm rad}}=S(x,y)-\frac{\delta h}{\tau_{rad}}.
\tag{A9}\label{A9}
\end{equation}
The quantity $S \equiv (h_{\rm eq}-H)/\tau_{\rm rad}$ represents the forcing. The radiative-equilibrium thickness is given by:
\begin{equation}
h_{\rm eq} = H + \Delta h_{\rm eq} \cos (k_x x)  \exp\left(-\frac{y^2}{2L_m^2}\right).
\tag{A10}  \label{A10}
\end{equation}
$\Delta h_{\rm eq}$ represents the day-night contrast in thickness at the substellar point when the system is in radiative equilibrium and the substellar point is at longitude 0\textdegree and latitude 0\textdegree. 
Based on the equilibrium thickness profile and assuming the forcing is symmetric about the equator, the first-order forcing term is $S\left(x,y\right) = \tau_{\rm rad}^{-1}\left(\Delta h_{\rm eq}/H\right) \exp\left(ik_x x\right) \exp\left(-y^2/2L_m^2\right)$. Set the forcing amplitude $\left(\Delta h_{eq}/H = 1\right)$.     

The induction equation takes the form:
\begin{equation}
\frac{\partial A}{\partial t}+V_x \frac{\partial A}{\partial x}+V_y\frac{\partial A}{\partial y}=\eta \Bigg[-\frac{1}{h}\frac{\partial h}{\partial x}\frac{\partial A}{\partial x}-\frac{1}{h}\frac{\partial h}{\partial y}\frac{\partial A}{\partial y}+\left(\frac{\partial^2}{\partial x^2}+\frac{\partial^2}{\partial y^2}\right)A\Bigg] .
\tag{A11}\label{A11}
\end{equation}
We again consider small perturbations to the physical quantities and linearize the equations:
\begin{equation}
\frac{\partial a}{\partial t}+u \frac{\partial A_0}{\partial x}+v\frac{\partial A_0}{\partial y}-\eta \Bigg[-\frac{1}{H}\frac{\partial \delta h}{\partial x}\frac{\partial A_0}{\partial x}-\frac{1}{H}\frac{\partial \delta h}{\partial y}\frac{\partial A_0}{\partial y}+\left(\frac{\partial^2}{\partial x^2}+\frac{\partial^2}{\partial y^2}\right)a\Bigg]=0 .
\tag{A12}\label{A12}
\end{equation}
Next, nondimensionalize equation \eqref{A8}, \eqref{A9} and \eqref{A12} with some dimensionless fluid numbers and consider steady-state flows to remove time-dependent term:
\begin{equation}
\begin{aligned}
g\frac{H}{L}\frac{\partial \delta h}{\partial x}-LU\beta yv-\frac{U^2}{L}\Bigg[-\left(\frac{\partial A_0}{\partial y}\right)^2\frac{\partial \delta h}{\partial x}+\frac{\partial A_0}{\partial y}\frac{\partial^2 a}{\partial x\partial y}-\frac{\partial^2A_0}{\partial y^2}\frac{\partial a}{\partial x}\Bigg]+\frac{U}{T}\frac{u}{\tau_{drag}}&=0,\\
g\frac{H}{L}\frac{\partial \delta h}{\partial y}+\beta LUyu+\frac{U^2}{L}\frac{\partial A_0}{\partial y}\frac{\partial^2a}{\partial x^2}+\frac{U}{T}\frac{v}{\tau_{drag}}&=0,\\
H\frac{U}{L}\left(\frac{\partial u}{\partial x}+\frac{\partial v}{\partial y}\right)-\frac{H}{T}S(x,y)+\frac{H}{T}\frac{\delta h}{\tau_{rad}}&=0,\\
HU^2v\frac{\partial A_0}{\partial y}-\eta \frac{HU}{L} \Bigg[-\frac{\partial \delta h}{\partial y}\frac{\partial A_0}{\partial y}+\left(\frac{\partial^2}{\partial x^2}+\frac{\partial^2}{\partial y^2}\right)a\Bigg]&=0 .
\end{aligned}
\tag{A13}\label{A13}
\end{equation}
Define the magnetic Reynolds number $R_B = \frac{UL}{\eta}$ to nondimensionalize further:
\begin{equation}
\begin{aligned}
\frac{\partial \delta h}{\partial x}-yv-\Bigg[-\left(\frac{\partial A_0}{\partial y}\right)^2\frac{\partial \delta h}{\partial x}+\frac{\partial A_0}{\partial y}\frac{\partial^2 a}{\partial x\partial y}-\frac{\partial^2A_0}{\partial y^2}\frac{\partial a}{\partial x}\Bigg]+\frac{u}{\tau_{drag}}&=0,\\
\frac{\partial \delta h}{\partial y}+yu+\frac{\partial A_0}{\partial y}\frac{\partial^2a}{\partial x^2}+\frac{v}{\tau_{drag}}&=0,\\
\left(\frac{\partial u}{\partial x}+\frac{\partial v}{\partial y}\right)-S(x,y)+\frac{\delta h}{\tau_{rad}}&=0,\\
v\frac{\partial A_0}{\partial y}- \frac{1}{R_B} \Bigg[-\frac{\partial \delta h}{\partial y}\frac{\partial A_0}{\partial y}+\left(\frac{\partial^2}{\partial x^2}+\frac{\partial^2}{\partial y^2}\right)a\Bigg]&=0.
\end{aligned}
\tag{A14}\label{A14}
\end{equation}

\section{zonal (east-west) acceleration of the zonal-mean flow }
The x- components of the momentum equation \eqref{A1} are:
\begin{equation}
\frac{\partial u}{\partial t}+u\frac{\partial u}{\partial x}+v\frac{\partial u}{\partial y}+g\frac{\partial h}{\partial x}-\beta yv-\left(B_x\frac{\partial}{\partial x}+B_y\frac{\partial}{\partial y}\right)B_x+\frac{Qu}{h}+\frac{u}{\tau_{drag}}= 0.
\tag{B1}\label{B1}
\end{equation}
Consider the zonal mean components and perturbations of the variables in equation \eqref{B1}:
\begin{equation}
\begin{aligned}
u&=\overline{u}+u',\\
v&=\overline{v}+v',\\
h  &=\overline{h}+h',\\
B_x&=\overline{b_x}+b'_x,\\
B_y&=\overline{b_y}+b'_y.
\end{aligned}
\tag{B2} \label{B2}
\end{equation}
Then, we introduce  the thickness-weighted zonal mean of any quantity $M$ 
$\left(\overline{M}^{*} \equiv \overline{hM}/\overline{h}\right)$ and derive separately the thickness-weighted zonal mean of each term in the equation (Note that $\overline{M'}=0$, $\overline{\partial_xM}=\frac{1}{L_x}\int_0^{L_x}\frac{\partial M}{\partial x}\delta x=\frac{M(L_x)-M(0)}{L_x}=0$, $L_x = \frac{2\pi}{k_x}$.):
\begin{equation}
\frac{1}{\overline{h}}\overline{h\frac{\partial V_x}{\partial t}}= \frac{1}{\overline{h}}\Bigg[\overline{h}\frac{\partial \overline{u}}{\partial t}-\overline{h'\frac{\partial u'}{\partial t}}\Bigg]=\frac{\partial \overline{u}}{\partial t}+\frac{1}{\overline{h}}\Bigg[\overline{\frac{\partial h'u'}{\partial t}}-\overline{u'\frac{\partial h'}{\partial t}}\Bigg],
\tag{B3} \label{B3} 
\end{equation}
\begin{equation}
\frac{1}{\overline{h}}\overline{\Bigg[hV_y\frac{\partial V_x}{\partial y}\Bigg]}=\frac{1}{\overline{h}} \Bigg[\overline{hV_y}\frac{\partial \overline{u}}{\partial y}+\overline{(hV_y)'\frac{\partial (\overline{u}+u')}{\partial y}}\Bigg]=\overline{V_y}^*\frac{\partial \overline{u}}{\partial y}+\frac{1}{\overline{h}}\frac{\partial}{\partial y}\left[\overline{(hv)'u'}\right] - \frac{1}{\overline{h}}\overline{u'\frac{\partial \left(hv\right)'}{\partial y}},
\tag{B4} \label{B4} 
\end{equation}

\begin{equation}
\frac{1}{\overline{h}}\overline{\Bigg[hB_y\frac{\partial B_x}{\partial y} \Bigg]} = \frac{1}{\overline{h}}\Bigg[\overline{\left(hB_y\right)}\frac{\partial \overline{b_x}}{\partial y}+\overline{\left(hB_y\right)'\frac{\partial b_x'} {\partial y}}\Bigg] = \overline{B_y}^*\frac{\partial \overline{b_x}}{\partial y}+\frac{1}{\overline{h}}\frac{\partial [\overline{(hB_y)'b_x'}]}{\partial y} - \frac{1}{\overline{h}}\overline{b_x'\frac{\partial \left(hB_y\right)'}{\partial y}}.
\tag{B5} \label{B5} 
\end{equation}

Multiply the continuity equation $\left(\frac{\partial h}{\partial t}+\frac{\partial\left(hu\right)}{\partial x}+\frac{\partial \left(hv\right)}{\partial y} = Q\right)$ by $u'$ and zonally average the equation:
\begin{equation}
\overline{u'\frac{\partial h'}{\partial t}}+\overline{u'\frac{\partial \left(hv\right)'}{\partial y}} = \overline{u'Q'}.
\tag{B6}\label{B6}
\end{equation}

Similarly, the SWMHD divergence-free condition $\left(\nabla \cdot h \textbf{B}=0\right)$ can be written as:
\begin{equation}
\frac{\partial \overline{hB_x}}{\partial x}+\frac{\partial \overline{hB_y}}{\partial y}=\frac{\partial \overline {hB_y}}{\partial y}=0,
\tag{B7}\label{B7}
\end{equation}
\begin{equation}
\overline{B_x\frac{\partial hB_x}{\partial x}}+\overline{B_x\frac{\partial hB_y}{\partial y}}=\overline{b_x}\frac{\partial \overline{hB_y}}{\partial y}+\frac{1}{\overline{h}}\overline{b_x'\frac{\partial \left(hB_y\right)'}{\partial y}}=\frac{1}{\overline{h}}\overline{b_x'\frac{\partial \left(hB_y\right)'}{\partial y}}=0.
\tag{B8}\label{B8}
\end{equation}

Rearranging equations \eqref{B3} - \eqref{B8} systematically gives the zonal acceleration equation of the zonal-mean flow:
\begin{equation}
\begin{aligned}
\frac{\partial \overline{u}}{\partial t} = 
\underbrace{ \overline{V_y}^{*}\Bigg[f-\frac{\partial \overline{u}}{\partial y}\Bigg]}_I -
\underbrace{ \frac{1}{\overline{h}}\frac{\partial}{\partial y}[\overline{(hv)'u'}]}_{II} +
\underbrace{\Bigg[\frac{1}{\overline{h}}\overline{u'Q'}+\overline{R_u}^{*}\Bigg]}_{III} \\-\underbrace{\frac{\overline{V_x}^{*}}{\tau_{drag}}}_{IV}+
\underbrace{\frac{1}{\overline{h}}\Bigg[\overline{\left(hB_y\right)}\frac{\partial \overline{b_x}}{\partial y}+\frac{1}{\overline{h}}\frac{\partial [\overline{(hB_y)'b_x'}]}{\partial y}\Bigg]} _{V}-
\frac{1}{\overline{h}}\frac{\partial(\overline{h'u'})}{\partial t}. 
\end{aligned}
\tag{B9}
\label{B9}
\end{equation}

\section{Pseudo-spectral method}
The pseudo-spectral method can transform the ordinary differential equations into the linearized equations. Consider the momentum perturbation equation in the $y$-direction:
\begin{equation}
    \frac{\partial \delta h}{\partial y}+yu+\frac{\partial A_0}{\partial y}\frac{\partial^2a}{\partial x^2}+\frac{v}{\tau_{drag}}=0.
    \tag{C1}
\label{C1}
\end{equation}
The solution is expanded as a sum of a series of basis functions:
\begin{equation}
\left\{u\left( y_i\right),v\left( y_i\right),\delta h\left( y_i\right),a\left( y_i\right)\right\}=\sum\limits_{j} \left\{\hat{u}_{ji}, \hat{v}_{ji}, \hat{\delta h}_{ji}, \hat{a}_{ji}\right\}\psi_j\left(y_i\right).
\tag{C2}
\label{C2}
\end{equation}
$y_i$ are the collocation points, $\psi_j$ is the basis function for every mode and $ \left\{\hat{u}_{ji}, \hat{v}_{ji}, \hat{\delta h}_{ji}, \hat{a}_{ji}\right\}$ is the coefficient.
The more collocation points and higher orders are chosen, the higher the accuracy becomes. We select parabolic cylinder functions as the basis. Parabolic cylinder functions can be simply expressed as the product of a Gaussian function and Hermite polynomials, with their collocation points corresponding to the zeros of the highest-order Hermite polynomial.Then we can get the equations:
\begin{equation}
\sum\limits_{j}\hat{\delta h}_{ji}\frac{d \psi_j\left(y_i\right)}{dy}+y\sum\limits_{j}\hat{u}_{ji}\psi_j\left(y_i\right)-\frac{\partial A_0}{\partial y}k_x^2\sum\limits_{j}\hat{a}_{ji}\psi_j\left(y_i\right)+\frac{1}{\tau_{drag}}\sum\limits_{j}\hat{v}_{ji}\psi_j\left(y_i\right)=0.
\tag{C3}
\label{C3}
\end{equation}
Similarly, the pseudo-spectral method can be applied to the other three equations in equation \eqref{A14} to calculate the corresponding coefficient for each variable, which are then substituted back into the basis functions to obtain the final solution:
\begin{equation}
\left\{u\left( x,y\right),v\left(  x,y\right),\delta h\left(  x,y\right),a\left(  x,y\right)\right\}=e^{ik_xx}\sum\limits_{n=0}^{N} \left\{\hat{u}, \hat{v} , \hat{\delta h} , \hat{a}\right\}\psi_n\left(y\right).
\tag{C4}
\label{C4}
\end{equation}

\bibliographystyle{raa}
\bibliography{ms2026-0235}

\end{document}